\PassOptionsToPackage{table,dvipsnames}{xcolor}
\documentclass[acmtog,nonacm]{acmart}
\setcopyright{none}
\usepackage{graphicx}
\usepackage{subcaption}
\usepackage{epigraph}

\usepackage{algorithm}

\usepackage{amssymb}

\newcommand{\ourmethod}{CoatFusion}
\newcommand{\ourdataset}{DataCoat110K}

\newcommand{\etal}{\emph{et al.}}

\begin{document}

\title[\ourmethod]{\ourmethod: Controllable Material Coating in Images}

\author{Sagie Levy}
\affiliation{%
  \institution{The Hebrew University of Jerusalem}
  \city{Jerusalem}
  \country{Israel}
}

\author{Elad Aharoni}
\affiliation{%
  \institution{The Hebrew University of Jerusalem}
  \city{Jerusalem}
  \country{Israel}
}

\author{Matan Levy}
\affiliation{%
  \institution{The Hebrew University of Jerusalem}
  \city{Jerusalem}
  \country{Israel}
}

\author{Ariel Shamir}
\affiliation{%
  \institution{Reichman University}
  \city{Herzliya}
  \country{Israel}
}

\author{Dani Lischinski}
\affiliation{%
  \institution{The Hebrew University of Jerusalem}
  \city{Jerusalem}
  \country{Israel}
}

\begin{teaserfigure}
    \centering\includegraphics[width=0.82\linewidth]{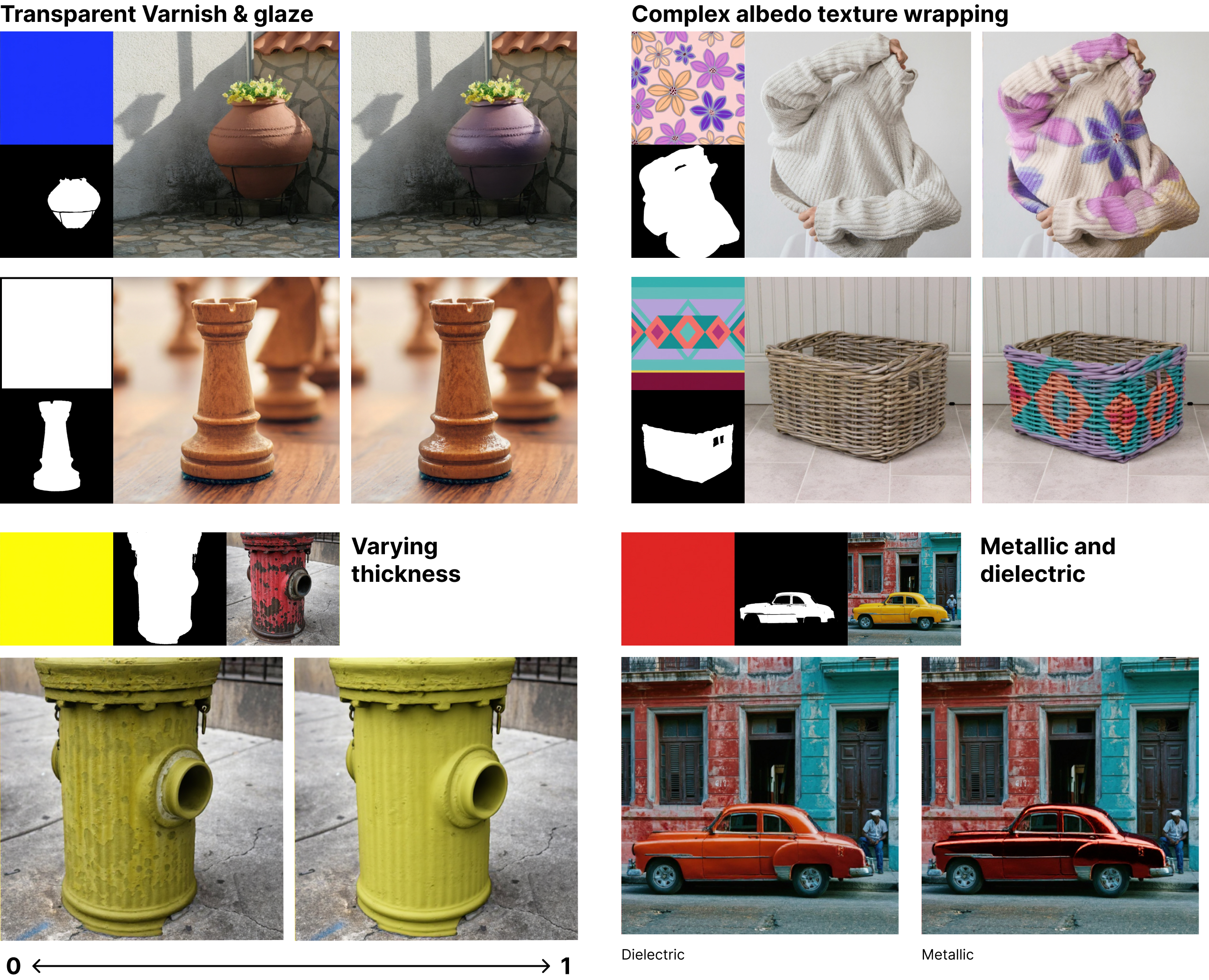}
\caption{
\label{fig:teaser}
\textbf{Overview.} Our method, \ourmethod{}, coats objects in input images with materials defined by albedo images, masks and parametric controls. \textbf{Top Left:} Objects coated with translucent materials: varnish and colorful glaze. \textbf{Top Right:} Objects with complex geometry covered by drawings and patterns, showing implicit understanding of texture warping. \textbf{Bottom Left:} Varying levels of coating material thickness. \textbf{Bottom Right:} Metallic and dielectric material coatings. \textbf{Note:} text labels are shown for the reader's convenience only and are not part of the model's input.
}
\Description{Four labelled groups of before-and-after photographs. Top left, translucent coatings: a terracotta hanging planter becomes glossy purple, and a matte wooden chess rook gains a wet-looking varnish; each is accompanied by its flat albedo swatch and a binary object mask. Top right, texture wrapping: a cream ribbed knit sweater held up by a person is covered with a large violet floral pattern that follows the folds of the fabric, and a grey wicker basket is covered with a teal and orange geometric pattern that follows the individual reeds of the weave. Bottom left, thickness: a red weathered fire hydrant coated in yellow at thickness zero, where the pitted and chipped surface still shows through, and at thickness one, where the surface is smooth. Bottom right, an orange vintage car on a Havana street coated as a matte dielectric and as a mirror-like metallic.}

\end{teaserfigure}

\begin{abstract}
    \emph{Material Coating} is a novel image editing task that simulates applying a thin material layer to a target surface (substrate) while preserving the substrate's geometry and shading. Material coating is fundamentally different from material transfer, which typically reinterprets an object as being made of a new material and can modify its shape or overwrite fine details. We introduce \emph{\ourmethod}, a diffusion-based framework for controllable material coating in images. Since real paired before/after coating data is difficult to acquire, we construct \emph{\ourdataset}, a synthetic paired dataset of approximately 110K examples rendered with physically motivated coating layers, including opaque, metallic, and transmissive appearances. \ourmethod{} formulates coating as visual in-context learning over a triptych containing the substrate image, a canonical 2D albedo texture, and the coated result, and uses cycle-consistent compose/decompose supervision to encourage substrate preservation and geometry-aware texture wrapping. Learned material tokens provide continuous control over roughness and thickness, and discrete on/off control over metalness and transmission. Across held-out synthetic evaluations, real-image examples, ablations, and a human study, \ourmethod{} produces more realistic and controllable coatings than material transfer/editing baselines and classical blending operations. Code and data for this paper are available at \url{https://github.com/sagielevy/coatFusion}.
    
\end{abstract}


\maketitle

\section{Introduction}

Editing the material properties of objects within images is a
long-standing challenge in computer graphics \cite{matusik2003reflectance,burley2012disneybsdf,sharma2023alchemistparametriccontrolmaterial,cheng2025marble,lopes2025matswap,garifullin2025materialfusion,Huang_2025_ICCV}. The ability to realistically alter an object's surface appearance has profound implications for a wide range of applications, including content creation, virtual prototyping, e-commerce visualization, and artistic expression. This editing task is most commonly framed as \emph{Material Transfer}, where the goal is to make an object appear as if it were intrinsically made from a different material~\cite{cheng2024zest, garifullin2025materialfusion,Huang_2025_ICCV}. However, this paradigm fails to address a common, yet fundamentally different, operation: \emph{Material Coating}.

\begin{figure}
    \centering
    \includegraphics[width=1.0\linewidth]{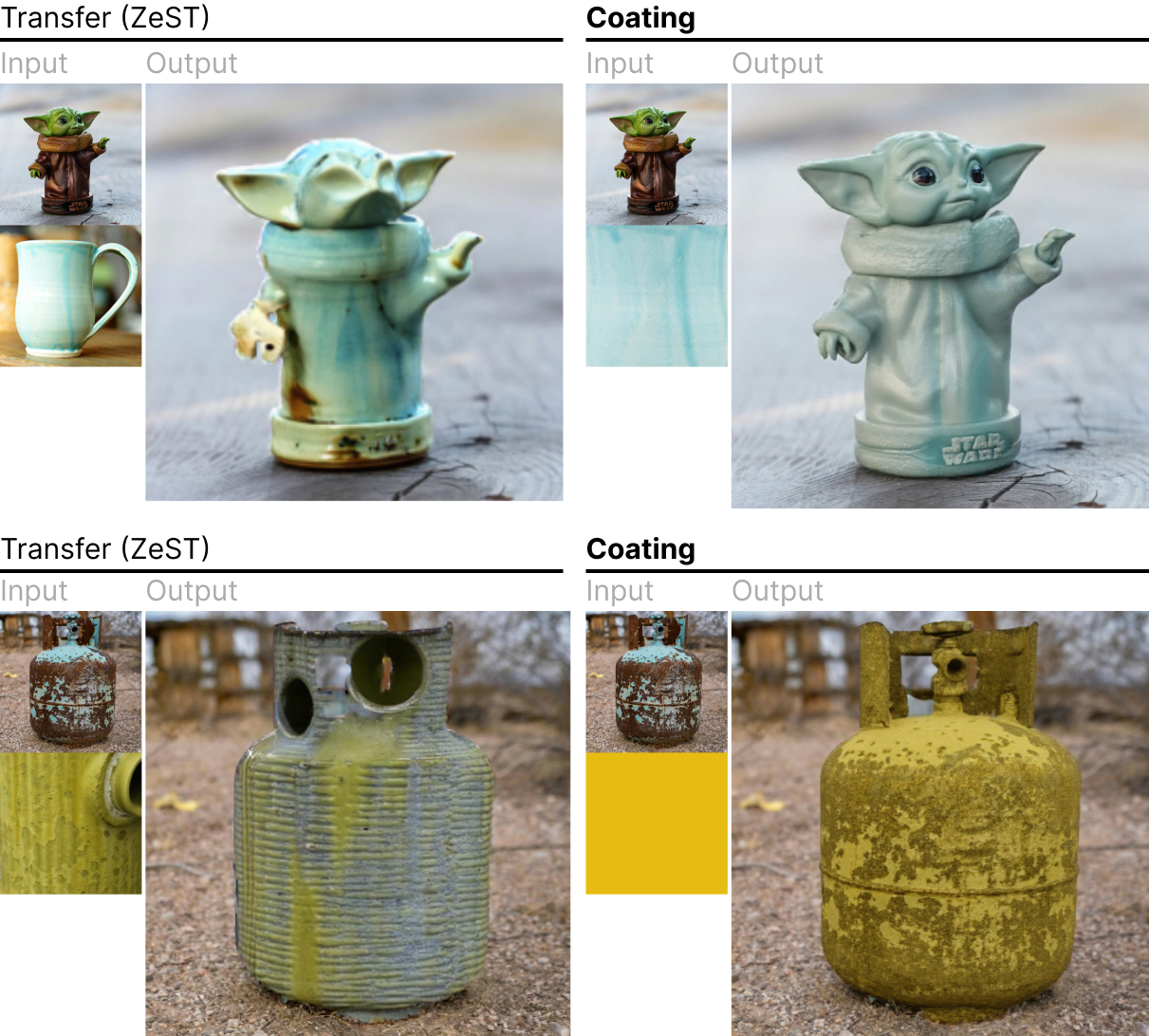}
    \caption{\textbf{Material Transfer vs.~Material Coating.} The latter task aims to maintain the geometry and fine-grained texture details of the original object, as if a layer of paint was applied on top of the geometry.}
    \Description{A two-by-two grid of photographs. Each column pair shows a small input photo and an exemplar swatch on the left and a large output on the right. Top row: a figurine of a large-eared, big-eyed green creature in a brown robe, with a mint-glazed ceramic mug as the exemplar. Material transfer by ZeST reinterprets the figurine as a glazed ceramic vessel, blurring the face and losing the folds of the robe; coating keeps the ears, the eyes, the folds of the robe and the embossed lettering on the base, tinting them mint green. Bottom row: a rusty teal propane cylinder with a plain yellow swatch. Transfer produces a ribbed ceramic pot that no longer resembles the cylinder; coating keeps the valve, the welded seam and the pitted rust while turning the surface yellow.}
    \label{fig:coating_vs_transfer}
\end{figure}

We define \emph{Material Coating} as the task of simulating a new, thin material layer \emph{on top of} a target object (the ``substrate''), such as applying a coat of paint, metallic spray, translucent glaze or a thin layer of some pattern. The distinction from material transfer is critical: a coating must preserve the substrate's underlying properties, most notably its geometry and fine-scale texture, which are precisely the properties that material transfer methods are designed to alter, as illustrated in Figure~\ref{fig:coating_vs_transfer}. Material coating unlocks common real-world editing scenarios that are inaccessible to transfer-based methods, such as applying varnish to furniture, adding graffiti to a textured brick wall, or visualizing new patterns on e-commerce products.

This fundamental need for substrate preservation exposes a critical gap in existing research. Recent state-of-the-art methods, such as
ZeST~\cite{cheng2024zest}, MARBLE~\cite{cheng2025marble}, 
MaterialFusion~\cite{garifullin2025materialfusion}, and  MaTe~\cite{Huang_2025_ICCV} are designed for exemplar-based
material transfer. Their goal is to make an object appear as if it were
made from the exemplar, which, by design, overwrites the original
object's fine details. Even light-aware methods like
MatSwap~\cite{lopes2025matswap}, or parametric editors like
Alchemist~\cite{sharma2023alchemistparametriccontrolmaterial}, which modify an object's \textit{existing}
properties (e.g., making it shinier), are not architecturally designed
to simulate the application of a new overcoat layer.

Despite the simplicity of its formulation, the Material Coating task remains particularly challenging for current learning-based methods. As shown in our evaluations, even non-learned, deterministic blending techniques from Photoshop, such as \textit{Blend If} and \textit{Color Blend Mode}, consistently outperform all existing Material Transfer approaches by a substantial margin in quantitative metrics. This observation underscores that, although closely related to Material Transfer, the Material Coating task is a distinct and unsolved problem that demands models explicitly designed to simulate physical over-coating rather than material replacement.

To address this new task, we introduce \textbf{\ourmethod}, a new approach for applying material coating in images. As real examples of images before/after coating are rare and difficult to create, we rely on a new large-scale synthetic dataset (110K images) that we specifically built for this problem, \ourdataset{}. We generate data by rendering a 3D object (the substrate) and then simulating a second, physically extruded ``cover'' mesh on top of it. This strategy, detailed in Section~\ref{sec:dataset}, explicitly teaches the network the appearance of \textit{coating} rather than \textit{replacing}.

Next, we propose a novel diffusion-based architecture \cite{Rombach22}, trained on this
dataset (Section~\ref{sec:method}). Our model is conditioned
\textit{simultaneously} on a 2D albedo texture, allowing
complex texture wrapping as shown in Figure~\ref{fig:teaser} (top right) and a
set of parametric granular, physically-based controls, as shown in the rest of Figure~\ref{fig:teaser}. These controls include roughness,
metalness, transmission and a key \textit{thickness} parameter. Thickness realistically modulates the coating's opacity and its interaction with the substrate's micro-geometry. 

To implement these capabilities, we reformulate material coating as a visual in-context learning task. Rather than relying on explicit geometric priors, the model is presented with a three-panel grid (triptych) that contains the substrate, the target albedo, and the desired coated result. When coating with an input pattern, the model also needs to learn a plausible geometric mapping onto the substrate. We utilize LoRA to allow the foundation model to learn parametric material control and either cubic or spherical mapping of flat textures onto 3D surfaces. To ensure the substrate's fine-grained details are preserved and the albedo is properly mapped to the geometry, we employ a cycle-consistent tuning strategy (Figure~\ref{fig:train_and_loss}). By jointly training coating and ``uncoating'' tasks, we enforce a bidirectional reconstruction consistency. This dual supervision enables the model to capture non-linear interactions such as complex geometry-aware wrapping.

Our qualitative and quantitative results (Section~\ref{sec:experiments}), later supported by a user study, 
demonstrate that \ourmethod{} produces realistic, controllable coatings
and significantly outperforms all material transfer baselines in this new task. We encourage the reader to watch the accompanying video for a comprehensive overview of our results.

Our contributions are threefold:
\begin{enumerate}
    \item We introduce \emph{Material Coating}, a new fine-grained image-editing task for altering the appearance of objects in images. 
    \item We introduce a large-scale, high-fidelity dataset and benchmark for the \emph{Material Coating} task.
    \item We propose \emph{CoatFusion}, a new diffusion-based approach for \emph{Material Coating} that provides fine-grained control via both 2D albedo maps and parametric attributes.
\end{enumerate}

\section{Related Work}

Diffusion models~\cite{ho2020denoising, ramesh2022hierarchical, Rombach22, Saharia2022PhotorealisticTD, Podell2023SDXLIL} have revolutionized high-quality image generation and editing. Recent models such as \textsc{Flux Fill}~\cite{fluxfill2025} support controllable image manipulation through mask conditioning and text-guided inpainting, enabling localized and semantically meaningful edits. Furthermore, recent works on image conditioning for Diffusion models, such as OminiControl2~\cite{tan2025ominicontrol2}, demonstrate the effectiveness of providing spatial conditioning alongside image inputs, providing strong priors for visual modifications and motivating alternative architectures like simple concatenation.

A major line of work in diffusion-based appearance editing focuses on \emph{material transfer}, where the goal is to make an object appear as if it were intrinsically composed of a different material. Exemplar-based approaches such as \textsc{ZeST}~\cite{cheng2024zest}, MaterialFusion~\cite{garifullin2025materialfusion}, MaTe~\cite{Huang_2025_ICCV}, MaterialPicker~\cite{Ma_2025}, and \textsc{MatSwap}~\cite{lopes2025matswap} transfer material appearance from reference images while attempting to preserve scene geometry and illumination cues. These methods employ increasingly sophisticated conditioning mechanisms, including ControlNet guidance, CLIP-based material representations, geometry-aware conditioning, and multimodal attention. Nevertheless, their primary objective remains \emph{material replacement}: the target object is reinterpreted as being made from the transferred material.

Another related direction focuses on \emph{parametric material editing}. Methods such as \textsc{Alchemist}~\cite{sharma2023alchemistparametriccontrolmaterial} and \textsc{MARBLE}~\cite{cheng2025marble}, or the Works by Delanoy \etal~\shortcite{delanoy2022generative} and Subias and Lagunas \shortcite{subias2023inthewild} provide explicit control over attributes such as \emph{roughness}, \emph{metalness}, \emph{transparency}, and \emph{albedo}. \textsc{MARBLE}, for example, performs material interpolation and blending in CLIP-space, while \textsc{Alchemist} directly manipulates material properties through synthetic supervision. While these methods provide fine-grained appearance control, they still operate under a material replacement paradigm rather than simulating the addition of a distinct coating layer on top of existing substrate geometry and appearance.

Despite the impressive realism achieved by these approaches, they remain unsuitable for the \emph{Material Coating} task considered in this work. Material coating requires preserving the substrate's underlying appearance, including its geometry, fine-scale texture, and shading structure, while simulating the application of a distinct overlaid layer. This distinction is particularly important for thin coatings such as paint, glaze, varnish, graffiti, or metallic spray, where the underlying substrate remains partially visible and continues to influence the final appearance. In contrast, material transfer methods often overwrite substrate details, suppress subtle geometric variations, or dominate the original reflectance entirely, as demonstrated in Figure~\ref{fig:coating_vs_transfer}.

Interestingly, we find that simple classical image editing heuristics can outperform modern learning-based material transfer methods on the coating task. In particular, Photoshop's \emph{Blend If} operation produces surprisingly effective coating-like results by preserving shadows and highlights from the original image while selectively blending the coating into mid-tone regions. However, it remains a hand-crafted operation without explicit control over coating properties such as \emph{roughness}, \emph{metalness}, \emph{transmission}, or \emph{thickness}.

In summary, while prior work has significantly advanced material transfer and material editing, no existing approach explicitly addresses \emph{Material Coating}. Our work fills this gap by introducing a coating-specific formulation, dataset, and diffusion-based architecture designed to preserve substrate appearance while enabling controllable coating properties such as roughness, metalness, transmission, and thickness.

\section{Dataset}
\label{sec:dataset}

Training a model for \emph{Material Coating} requires paired images showing the same object before and after the application of a coating layer. Since no such large-scale dataset exists, and physically capturing paired data with controlled material attributes is impractical, we construct a synthetic dataset, \ourdataset{}, using Blender~\shortcite{blender}.

\ourdataset{} is designed to support coating \emph{addition} and \emph{removal} while covering a broad range of objects, scenes, lighting conditions, and coating parameters. We render 104 unique object meshes for training, for a total of 1560 scenes, each observed from 15 camera viewpoints under one of 400 environment maps, 20 floor materials, and two randomly positioned light sources (assets and their sources are listed in the Appendix). All images were rendered using 200 samples per pixel (SPP). For every scene, we render one variant with the uncoated object and 64 coated variants generated with different coating parameters. A consistent random mask is projected from the camera onto the object for all variants, enabling controlled supervision for coating operations.

\begin{figure}[t]
    \centering
    \includegraphics[width=1.0\linewidth]{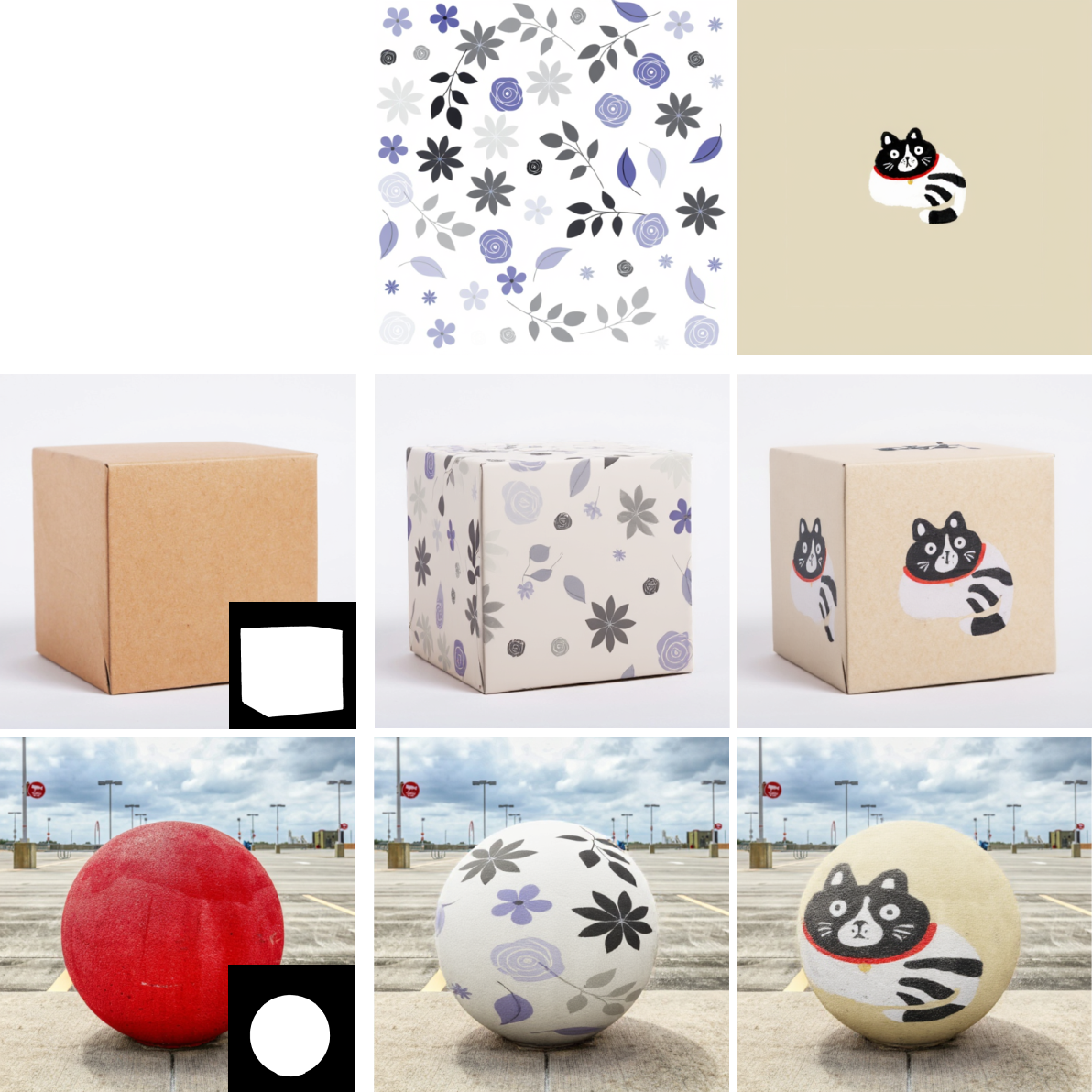}
    \caption{\textbf{Implicit geometry-aware texture wrapping}. The model learns to map canonical albedo textures onto complex object geometries.}
    \Description{Two rows of three photographs, headed by two flat albedo textures: a violet and grey floral pattern and a cartoon black-and-white cat on a beige ground. Top row: a plain cardboard box, with its binary mask inset, then the same box printed with the floral pattern continuing correctly across the top, front and side faces, then the same box with the cat rendered once across the front and wrapping around the corner onto the side. Bottom row: a red rubber ball in a parking lot, with its mask inset, then the same ball with the floral pattern curving over the sphere, then with the cat stretched around its curvature.}
    \label{fig:uv_mapping_comparison}
\end{figure}

The coating parameters include \emph{roughness}, \emph{metalness}, \emph{transmission}, \emph{thickness}, and \emph{albedo}.
The albedo is either a uniform color or a 2D pattern containing graphics, logos, graffiti, or repeating patterns. To apply these patterns as coating onto the objects, we use either cube projection, spherical projection, or the object's native UV map where appropriate. This allows the model to learn geometry-aware wrapping of the pattern for coating (Figure~\ref{fig:uv_mapping_comparison}).

\begin{figure}[t]
    \centering
    \includegraphics[width=1.0\linewidth]{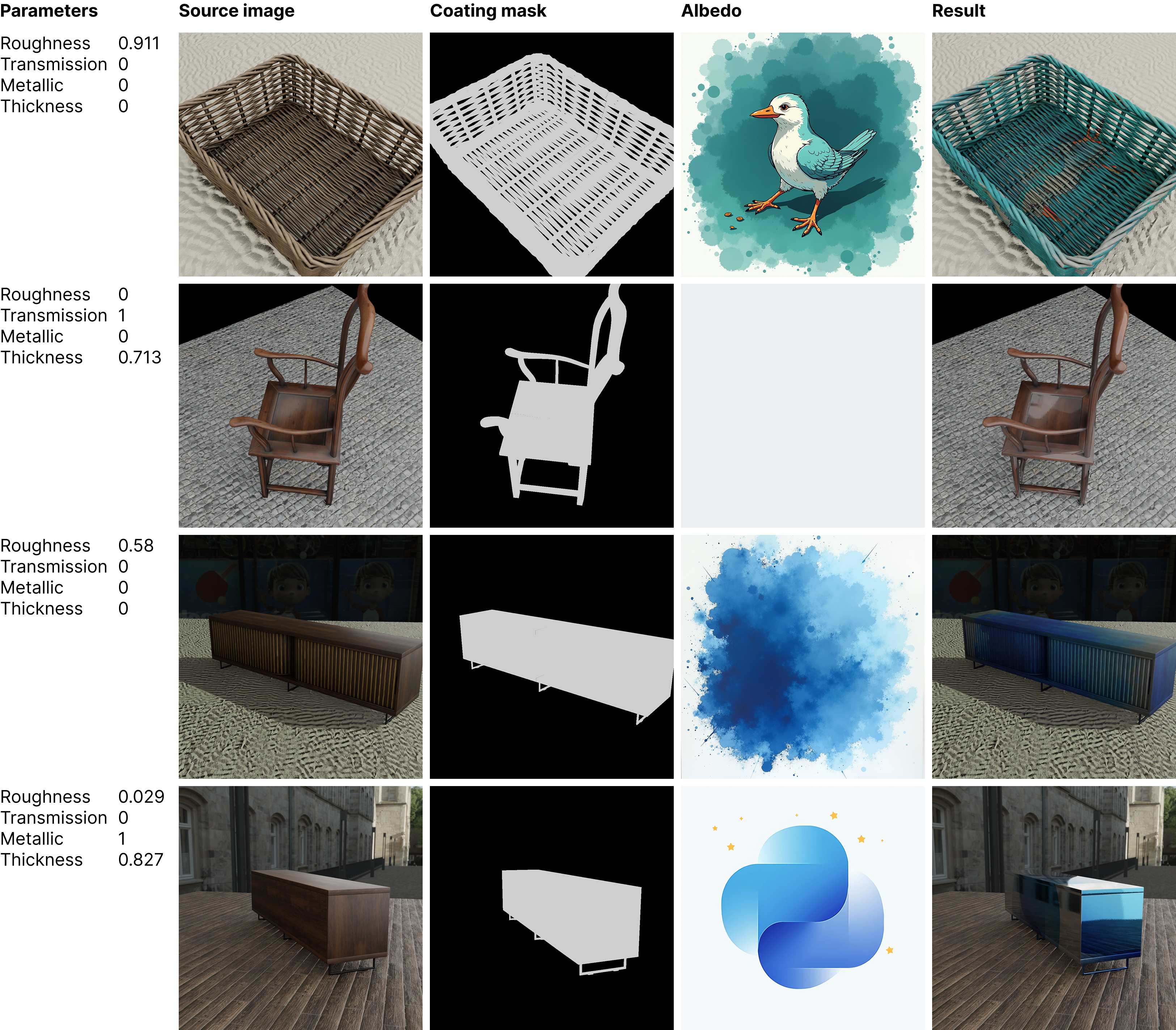}
    \caption{Samples from the held-out test set, used as our benchmark dataset.}
    \Description{A table of four rendered examples. Columns are the coating parameters, the source image, the binary coating mask, the flat albedo texture, and the resulting coated render. Rows: a woven wicker tray coated with a teal bird illustration; a wooden armchair given a transmissive coating at thickness 0.713 that softens its outline; a slatted media console coated with a blue watercolor wash; and a wooden sideboard given a highly metallic coating carrying a blue swirl logo.}
    \label{fig:benchmark_set}
\end{figure}

\begin{figure}[ht]
    \centering
    \includegraphics[width=1.0\linewidth]{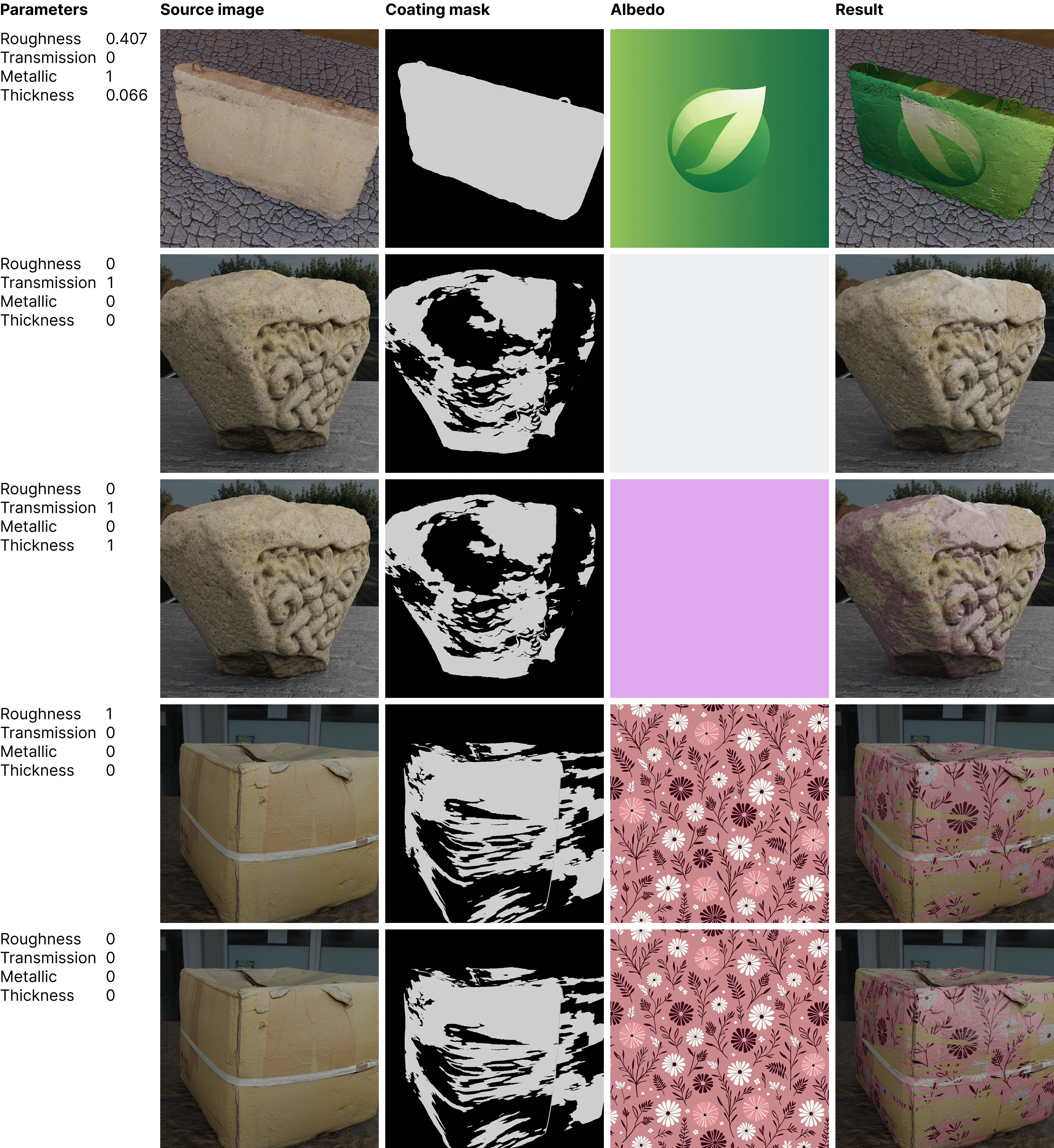}
    \caption{Examples from the dataset generation pipeline, generated in Blender. Each row shows (from left to right) the coating parameters, the original image, the coating mask, the albedo texture, and the resulting image.}
    \Description{A table of five rendered examples with columns for the coating parameters, source image, binary coating mask, albedo texture, and result. Rows: a pale stone bench coated with a metallic green leaf logo; a carved stone capital with a transmissive coating at thickness zero, which leaves it nearly unchanged; the same capital at thickness one, now tinted pink; a cardboard bale coated with a pink daisy pattern at roughness one, giving a matte finish; and the same bale at roughness zero, giving a glossier one. The masks are irregular, covering only part of each object.}
    \label{fig:dataset}
\end{figure}

Overall, \ourdataset{} contains approximately 110K paired images (see Figure~\ref{fig:dataset}). We also construct a held-out test split of approximately 10K examples using 12 unseen object meshes, textures, environment maps, and lighting configurations (a sample from the benchmark dataset is shown in Figure~\ref{fig:benchmark_set}).

\subsection{Rendering Coating}
\label{sec:render_coating}

To generate realistic coatings, we model the coating as a separate volumetric shell rather than relying on standard single-surface physically-based rendering (PBR) approximations such as Clearcoat \cite{burley2012disneybsdf}. Specifically, we duplicate the target mesh and extrude it slightly along its normals, producing a thin outer shell surrounding the original geometry.

This layered representation enables two important effects. First, it supports the full range of coating appearances, from opaque paints to transmissive varnishes. Second, it enables volumetric light attenuation inside the coating shell using Beer-Lambert absorption \cite{Kajiya1984}, producing thickness-dependent transmissive effects that cannot be reproduced with single surface shader mixing.

\begin{figure}[!ht]
    \centering
    \includegraphics[width=\linewidth]{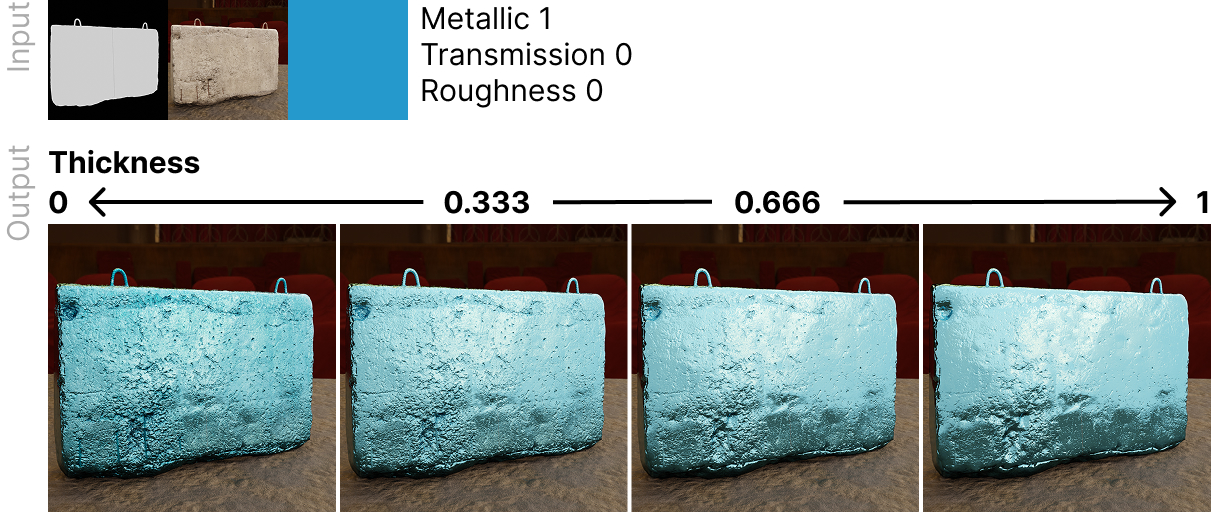}
    \caption{\textbf{Thickness variation.} A sample visualization showing the variation of thickness present in our dataset.}
   \Description{A row of inputs above a row of four renders. The inputs are a binary mask, a weathered stone slab, a solid cyan albedo, and the settings metallic one, transmission zero, roughness zero. The four renders show thickness increasing from zero to one along an arrow: at low thickness the cyan metallic coat follows every pit and crack of the stone, and by thickness one the surface is a smooth cyan shell in which the substrate relief is largely flattened out.}
   \label{fig:thickness_variation_dataset}
\end{figure}

While properties such as roughness and metalness are straightforward to parameterize~\cite{burley2012disneybsdf}, coating \emph{thickness} is more challenging because its visual effect depends strongly on the optical behavior of the coating material. To model this in a perceptually plausible way (see Figure~\ref{fig:thickness_variation_dataset} for a sample visualization of thickness variation), we introduce a virtual thickness parameter that jointly controls:
\begin{itemize}
    \item Attenuation of substrate normals,
    \item Volumetric absorption for transmissive coatings,
    \item Albedo blending for opaque coatings.
\end{itemize}

As thickness increases, the coating progressively smooths the substrate's microstructure by reducing the contribution of the original normal map.

For transmissive coatings, we utilize Blender's Volume Absorption node along with a fully transmissive BSDF shader. The Volume Absorption node's density factor is scaled according to thickness, producing increasingly opaque appearances.

For opaque coatings, we blend the coating and substrate albedos in OkLAB space~\cite{ottosson2020oklab}, preserving darker substrate details beneath thin coatings:
\[
\texttt{lerp(min(}\Theta(o).x,\Theta(c).x\texttt{),}\Theta(c).x, T\texttt{)}
\]
where $\Theta$ denotes the RGB-to-OkLAB transform, $o$ is the substrate albedo, $c$ is the coating albedo, and $T$ is the mapped thickness value (\textit{Opaque Albedo Blending} for the input thickness value $t$). This produces the appearance of a thin layer partially revealing the underlying surface texture, such as scratches or small geometric variations.

\begin{figure}[htbp]
    \centering
    \includegraphics[width=\linewidth]{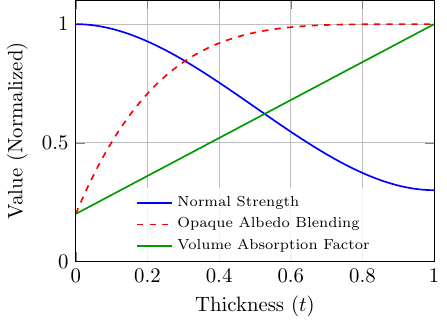}
    \caption{Thickness-dependent mappings of rendering parameters. Increasing thickness reduces \textbf{Normal Strength} non-linearly, suppressing substrate microstructure and producing a smoother coated surface; increases the \textbf{Volume Absorption Factor} for transmissive coatings, making them less transparent; and raises the \textbf{Opaque Albedo Blending} weight, moving opaque coatings non-linearly toward the coating albedo while preserving more of the substrate albedo at low thickness values. Formulas are detailed in the Appendix.}
    \Description{A line plot with thickness t from 0 to 1 on the horizontal axis and normalized value from 0 to 1 on the vertical axis. A blue solid curve, Normal Strength, stays near 1 until about t equals 0.3 and then falls in an S shape to about 0.3 at t equals 1. A red dashed curve, Opaque Albedo Blending, rises steeply from about 0.2 and saturates near 1 by t equals 0.6. A green solid line, Volume Absorption Factor, rises roughly linearly from about 0.2 to 1.}
    \label{fig:thickness_functions}
\end{figure}

\begin{figure}
    \centering
    \includegraphics[width=0.93\linewidth]{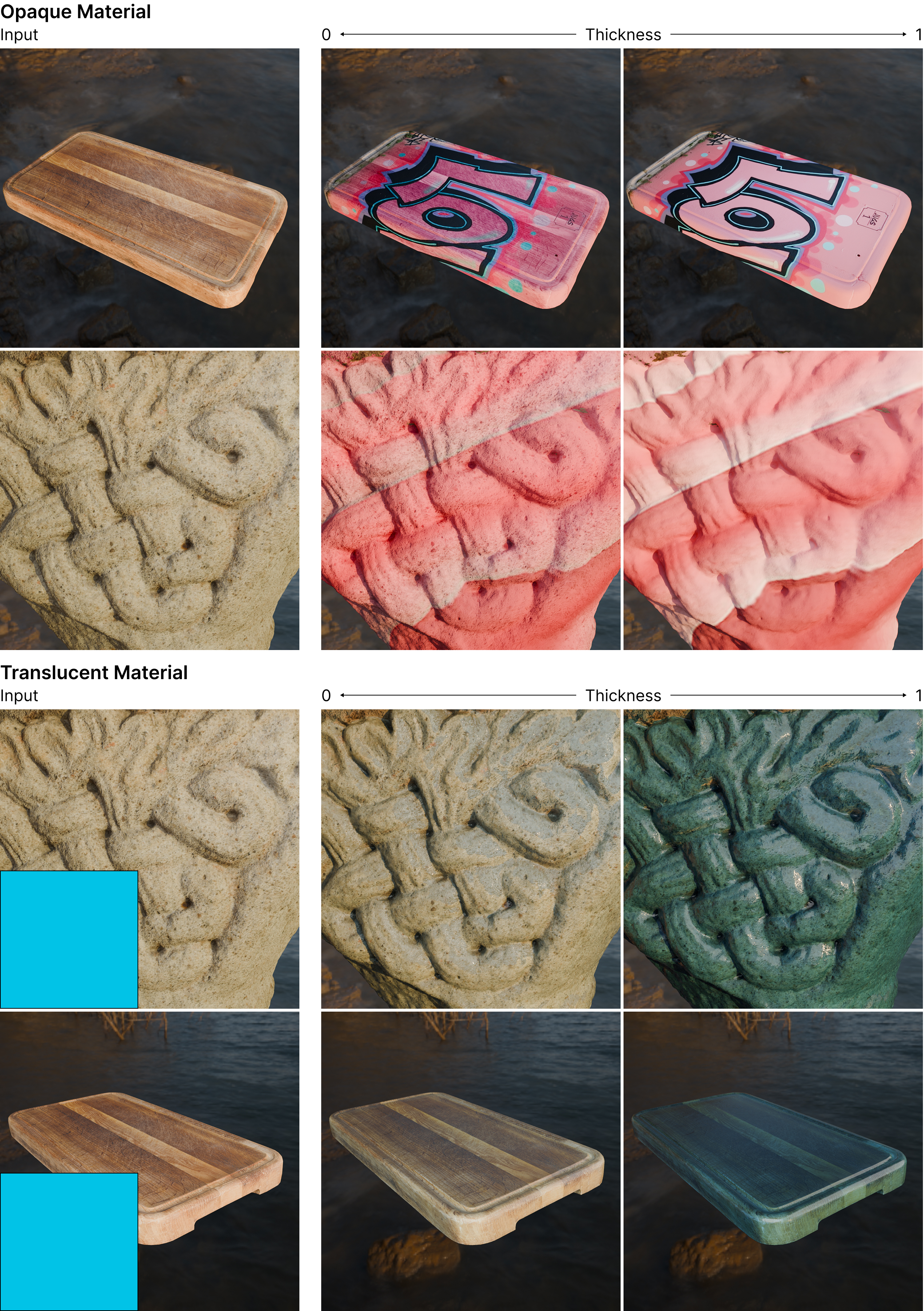}
    \caption{\textbf{Varying thickness}: Examples of applying a coating with minimal and maximal thickness levels. Note the increase in smoothness for the opaque coating (top rows), and the increasing visibility of the bluish coating color for the transmissive coating (bottom rows).}
    \Description{Two labelled blocks, each showing an input beside results at thickness zero and thickness one. In the opaque block, a wooden cutting board floating over water receives a pink graffiti coating whose painted numeral stays crisp while the wood grain disappears at high thickness, and a carved stone knotwork relief coated in pink keeps its deep carving at thickness zero but flattens into smooth pink at thickness one. In the translucent block, a cyan albedo swatch is applied to the same stone knotwork and to the wooden board: at thickness zero the coating is nearly invisible, and at thickness one both objects take on a deep blue-green tint through which the carving and the grain remain visible.}
    \label{fig:thickness_examples}
\end{figure}

Figure~\ref{fig:thickness_functions} visualizes the mapping between thickness and the different coating properties, while Figure~\ref{fig:thickness_examples} demonstrates the resulting appearance variations.

\section{Method}
\label{sec:method}

\begin{figure}
    \centering
    \includegraphics[width=\linewidth]{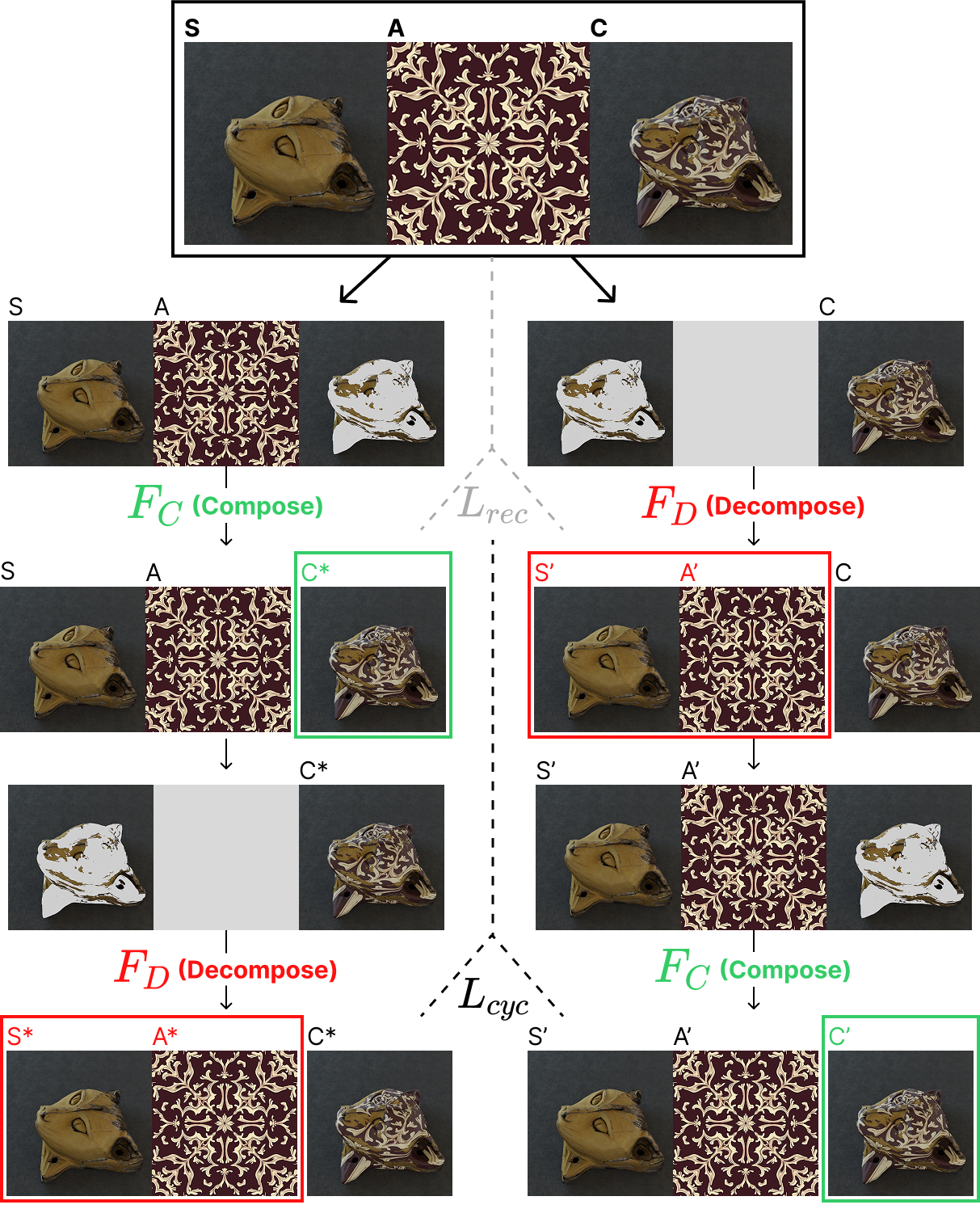}
    \caption{\textbf{Training and losses.} The model learns to compose and decompose the coating on the substrate. Training steps alternate between compose and decompose tracks. For the compose track (left): we first mask and predict the coating $C^*$. Next, we mask the substrate $S$ and albedo $A$ and predict them using $C^*$. The decompose track (right) runs in the reverse order. See Section~\ref{cycle_supervision} for details. Note that the mask channel is omitted to save space (instead, masked areas are filled with gray on the input triptychs).}
    \Description{A flow diagram over an example whose substrate S is a carved animal skull, whose albedo A is a dark red and cream damask pattern, and whose coated result C is the skull covered in that damask. The complete triptych at the top splits into two branches. On the left, the compose track: the C panel is masked out in grey, the operator F sub C predicts it, and the prediction C star is outlined in green; the result is then fed back with S and A masked, and F sub D recovers them as S star and A star, outlined in red. On the right, the decompose track runs the same steps in the opposite order, producing S prime and A prime outlined in red and then C prime outlined in green. A reconstruction loss L sub rec links the two first-stage predictions and a cycle loss L sub cyc links the two second-stage predictions.}
    \label{fig:train_and_loss}
\end{figure}

\ourmethod{} is a diffusion-based framework for \emph{Material Coating} that combines a triptych-based visual in-context learning (VICL) formulation with cycle-consistent supervision.

We build upon the cycle-consistent tuning approach introduced by Gu \etal~\shortcite{gu2026cycle} for layered image decomposition, extending it from image-space layer separation to geometry-aware material coating. Unlike prior decomposition work, our objective is not decomposition itself, but rather using decomposition as an auxiliary constraint that improves coating fidelity and texture-to-geometry correspondence.

\subsection{Visual In-Context Learning via Triptychs}

Rather than conditioning the diffusion model through latent concatenation alone (see Figure~\ref{fig:ablation_study}), we formulate material coating as a visual in-context learning task using a three-panel grid image (triptych), as illustrated in Figure~\ref{fig:train_and_loss}. The triptych $I$ is defined as:
\[
I = [S, A, C]
\]
where:
\begin{itemize}
    \item $S$ (\textbf{Substrate}) is the original image containing the uncoated object,
    \item $A$ (\textbf{Albedo}) is the coating albedo 2D texture in a canonical, unwrapped form,
    \item $C$ (\textbf{Coating}) is the output image containing the coated object.
\end{itemize}

Specifically, the triptych is formed by spatially concatenating the substrate image $S$, the albedo texture $A$, and the coated image $C$, and passed to the VAE as a single image. The diffusion model receives the triptych as input together with a binary mask $M$ (in a separate channel) defining the target region for inpainting (ROI).

For the \textbf{composition task}, the model receives $S$ and $A$ as context while the masked ROI in $C$ is removed and synthesized by the model. Unlike Gu \etal~\shortcite{gu2026cycle}, who mask the entire output panel, we only mask the user-specified coating region. This allows localized coating edits while preserving the surrounding image content.

For the auxiliary \textbf{decomposition task}, the unmasked coated image $C$ is provided as context while the corresponding region in $S$ is masked and $A$ is fully masked. Both are reconstructed by the model. This encourages the network to recover canonical albedo representations that remain consistent with the coated appearance.

We train only a Low-Rank Adaptation (LoRA) with rank 128, keeping the \textsc{Flux Fill} backbone frozen. This allows efficient specialization to coating-specific appearance transformations while preserving the pretrained model's broader image priors.

\subsection{Global Material Trait Conditioning}

While the triptych provides spatial and textural context, we use additional explicit fine-grained control over coating appearance. We introduce learned global conditioning tokens for \emph{roughness}, \emph{metalness}, \emph{transmission}, and \emph{thickness}.

\begin{figure}
    \centering
    \includegraphics[width=1.0\linewidth]{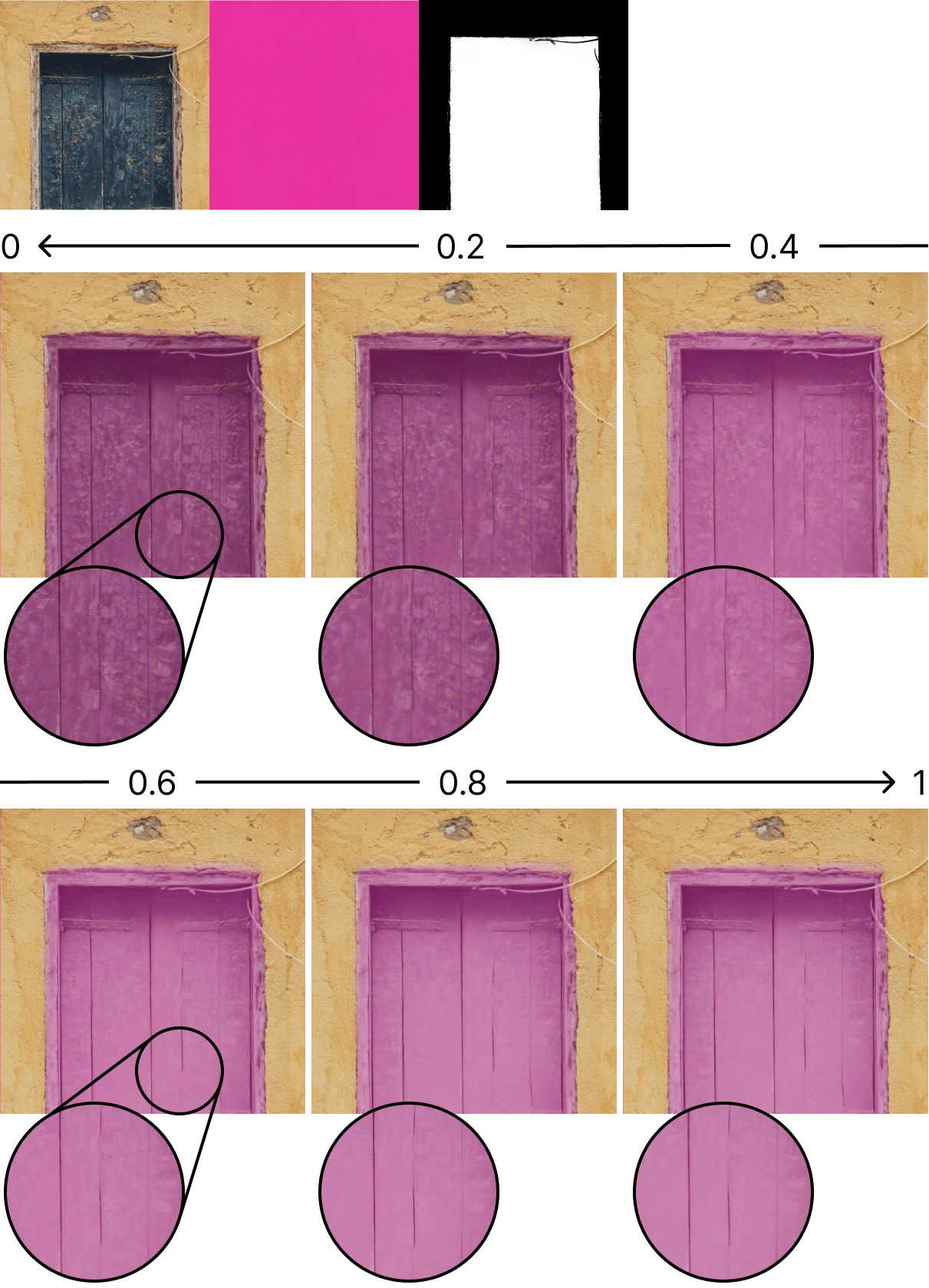}
    \caption{\textbf{Thickness comparison}. Increasing thickness progressively smooths substrate details and reduces transmissivity.}
    \Description{Inputs at the top: a photograph of a weathered blue-grey wooden door set in an ochre plaster wall, a solid magenta albedo, and a binary mask covering the door. Below, six results at thickness 0, 0.2, 0.4, 0.6, 0.8 and 1, each with a circular zoom inset on the door panel. At low thickness the pink coating still reveals the flaking paint, wood grain and nail marks; as thickness increases these fade until the door reads as a smooth, uniformly pink surface with only the plank seams remaining.}
    \label{fig:thickness_comparison}
\end{figure}

\begin{figure}
    \centering
    \includegraphics[width=1.0\linewidth]{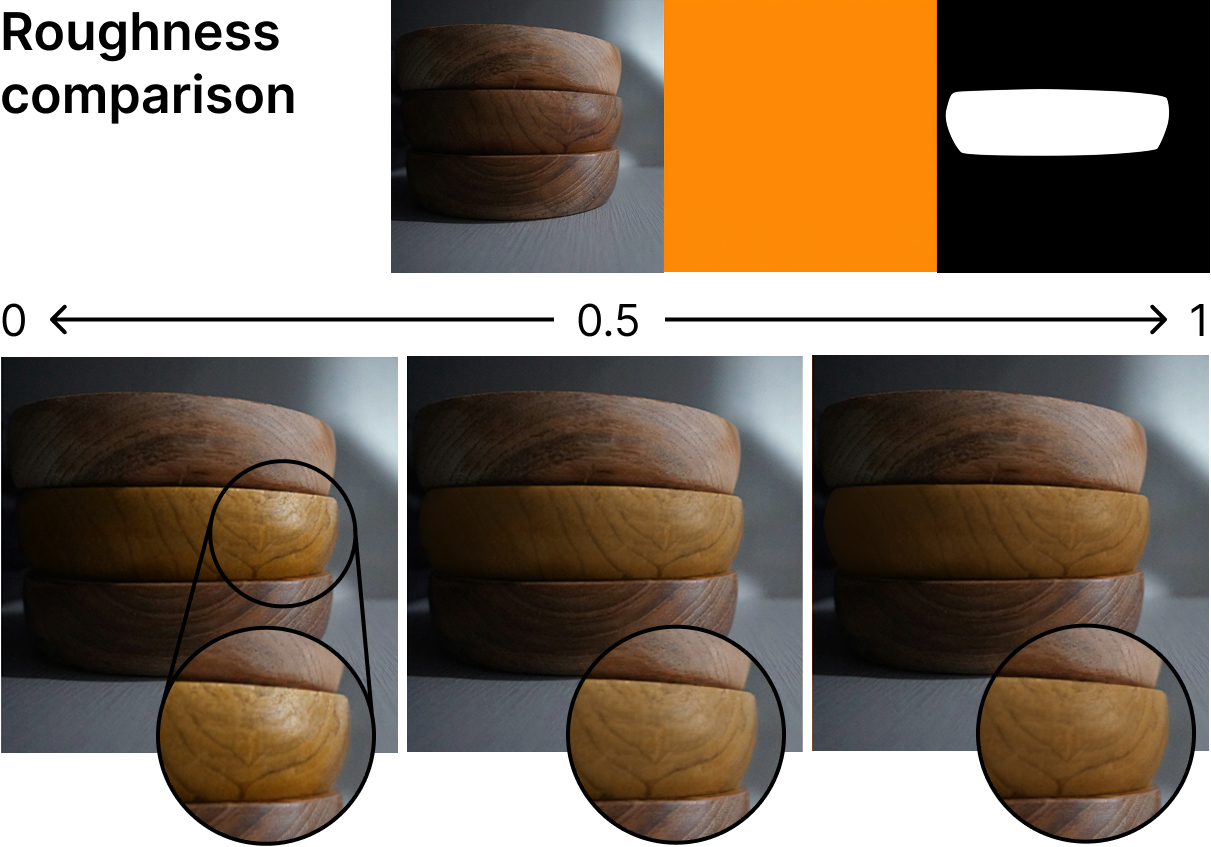}
    \caption{\textbf{Roughness control}. The coating appearance changes continuously with the roughness parameter while preserving the underlying geometry.}
    \Description{Inputs at the top: a stack of three turned wooden bowls in raking light, a solid orange albedo, and a mask covering only the middle bowl. Below, three results at roughness 0, 0.5 and 1, each with a circular zoom inset. At roughness zero the coated middle bowl carries a broad specular highlight and looks lacquered; at roughness one it is matte and diffuse. The upper and lower bowls and the cast shadows are unchanged throughout.}
    \label{fig:roughness_comparison}
\end{figure}

\emph{Roughness} and \emph{thickness} are continuous traits: each is controlled by a scalar value $\alpha \in [0,1]$ and represented using a composite embedding:
\[
E = E_{\text{trait}} + \alpha \cdot E_{\text{val}}
\]
where $E_{\text{trait}}$ identifies the attribute type and $E_{\text{val}}$ encodes its magnitude, providing continuous control over the generated coating (see examples in Figures~\ref{fig:teaser}, \ref{fig:thickness_comparison}, and \ref{fig:roughness_comparison}).

\emph{Metalness} and \emph{transmission}, in contrast, are treated as discrete traits: rather than a continuous embedding, each is represented by a dedicated learned ``on/off'' embedding (e.g., metallic vs.\ dielectric, transmissive vs.\ opaque).


\subsection{Cycle-Consistent Supervision}
\label{cycle_supervision}

The mapping between a canonical albedo texture and a 3D object appearance is inherently ambiguous due to perspective distortion, shading, and geometry-dependent texture deformation. To regularize this problem, we employ cycle-consistent supervision inspired by Gu \etal~\shortcite{gu2026cycle}.

We jointly optimize 1.\ a composition function $\mathcal{F}_C$ that applies a coating, and 2.\ an auxiliary decomposition function $\mathcal{F}_D$ that recovers the substrate and canonical albedo.

Training alternates between two complementary tracks:

\begin{center}
    \small
    \begin{tabular}{@{}lll@{}}
        \toprule
        \textbf{Phase} & \textbf{Track 1: Comp.-first} & \textbf{Track 2: Decomp.-first} \\
        \midrule
        Phase 1 & $C^* = \mathcal{F}_C(\langle S, A \rangle)$ & $\langle S', A' \rangle = \mathcal{F}_D(C)$ \\
        Phase 2 & $\langle S^*, A^* \rangle = \mathcal{F}_D(C^*)$ & $C' = \mathcal{F}_C(\langle S', A' \rangle)$ \\
        \bottomrule
    \end{tabular}
\end{center}

This cyclic supervision encourages consistency between coated appearances and their underlying canonical albedo representations. In practice, due to memory constraints, we evaluate only one track per training step and alternate between them during optimization (see Figure~\ref{fig:train_and_loss}).

\subsection{Training Objective}

We optimize the model using a standard flow-matching objective \cite{lipman2022flowmatching} combined with a cycle-consistency loss~\cite{zhu2017unpaired, gu2026cycle}. The total loss is defined as $\mathcal{L} = \mathcal{L}_{\text{rec}} + \mathcal{L}_{\text{cyc}}$. 

The reconstruction term $\mathcal{L}_{\text{rec}}$ is the standard rectified flow loss. Given clean latent $x_0$, noise $\epsilon \sim \mathcal{N}(0,I)$, interpolant $x_t = (1-t)x_0 + t\epsilon$, and network $v_\theta(x_t, t, c)$ predicting target velocity $\epsilon - x_0$, the loss at timestep $t_1$ is:
\begin{equation}
\mathcal{L}_{\text{rec}} = \mathbb{E}_{x_0,\epsilon,t_1} \big\lVert v_\theta(x_{t_1}, t_1, c) - (\epsilon - x_0) \big\rVert_2^2
\end{equation}
where $c$ is the task conditioning.

For cycle-consistency, evaluated at an independent timestep $t_2$, the \textbf{compose-first} track first estimates the clean output $\hat{x}_0 = x_{t_1} - t_1 v_\theta(x_{t_1}, t_1, c)$.
The predicted reconstruction is then re-noised using a fresh Gaussian sample $\epsilon' \sim \mathcal{N}(0,I)$ to obtain $\hat{x}_{t_2} = (1-t_2)\hat{x}_0 + t_2\epsilon'$.
For the second stage, the conditioning is switched to the complementary (decomposition) task by providing the coated image panel, $c'=\langle C\rangle$. The cycle-consistency loss is defined as
\begin{equation}\mathcal{L}_{\text{cyc}} =\mathbb{E}_{\epsilon',t_2}\left\|v_\theta\!\left(\hat{x}_{t_2}, t_2, c'\right)-(\epsilon' - x_0)\right\|_2^2 \end{equation} The \textbf{decompose-first} track is defined analogously, using $c'=\langle S, A\rangle$.

\subsection{Training Details}
\label{sec:training_details}

The training leverages a low-rank adaptation (LoRA) of rank 128 while keeping the \textsc{Flux Fill} backbone frozen. We optimize the model using AdamW with a constant learning rate of $1\times 10^{-4}$. The initial stage runs for 30,000 steps at $256 \times 256$ resolution on 3 H200 GPUs, followed by a second stage of 5,000 steps at $512 \times 512$ resolution on 4 H200 GPUs. For more implementation and training details, see the Appendix.

\section{Experiments}
\label{sec:experiments}
In this section, we evaluate our method against several baselines on the Material Coating task, followed by an ablation study on our design choice (VICL with cyclic consistency), and a human evaluation to assess the perceptual realism of our results. 

\begin{figure*}[ht]
    \centering
    \includegraphics[width=\linewidth]{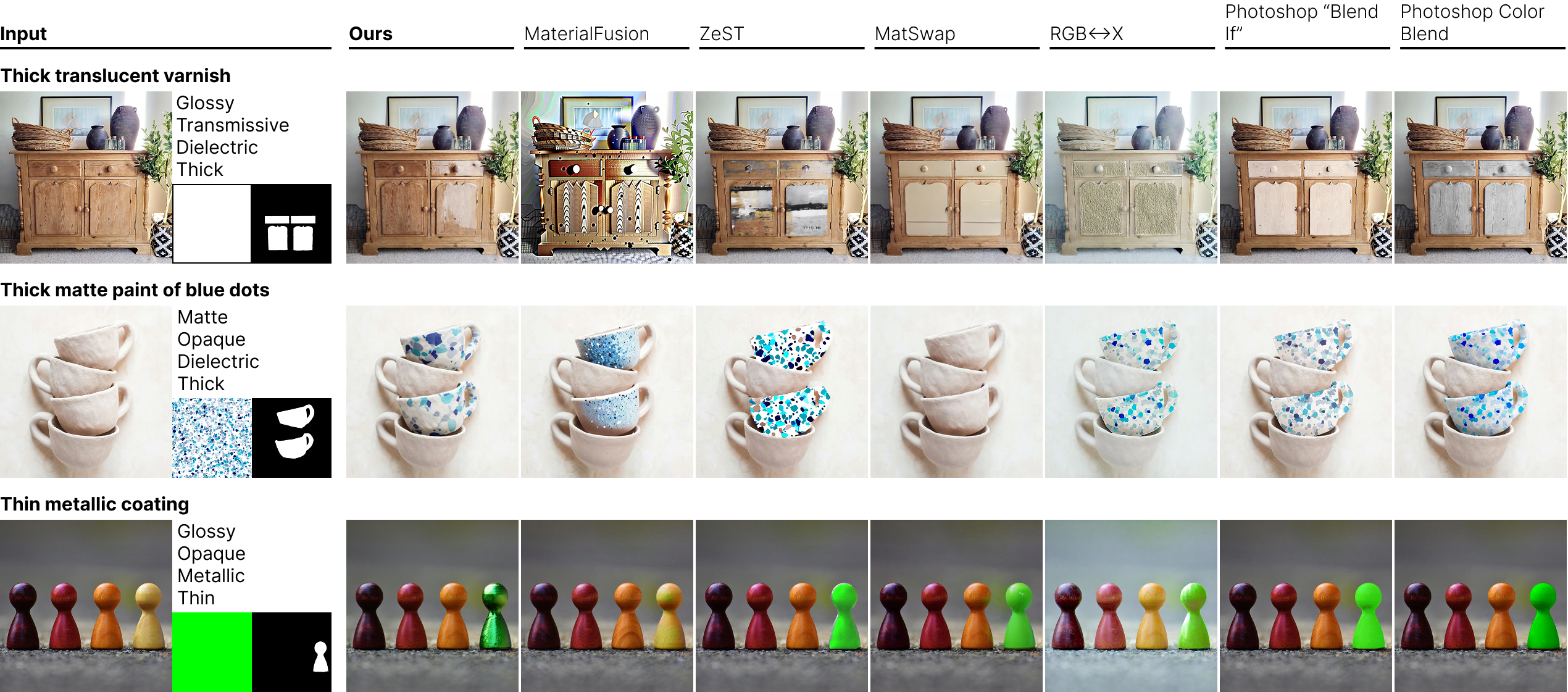}
    \caption{Comparison of \ourmethod{} with baselines on the material coating task. From top to bottom: applying thick translucent varnish to wooden cabinet doors, thick matte paint with a dotted pattern on clay mugs, thin metallic coating on the rightmost figure. Our method best adheres to the specified material characteristics, while preserving the fine details of the underlying geometry.}
    \Description{A grid with three examples in rows and eight columns: the input, our result, and results from MaterialFusion, ZeST, MatSwap, RGB to X, Photoshop Blend If, and Photoshop Color Blend. The input cell of each row holds the source photograph, the requested material traits as text, the albedo swatch and the mask. Row one applies a thick translucent varnish to the doors of a wooden sideboard; the transfer baselines repaint or restyle the cabinet, while ours darkens and glosses the doors only. Row two applies a thick matte blue dotted paint to a stack of ceramic cups; several baselines scatter the dots over the whole stack or ignore them, while ours confines them to the cups and follows their curvature. Row three applies a thin green metallic coating to the rightmost of a row of wooden pawns; most baselines recolor several pawns or lose the wood grain.}
    \label{fig:comparisons}
\end{figure*}

\subsection{Comparisons}
\label{sec:comparisons}

We evaluate \ourmethod{} qualitatively (Figure~\ref{fig:comparisons}) and quantitatively (Table~\ref{tab:metrics_results}) against several diffusion-based material transfer and editing methods~\cite{garifullin2025materialfusion, cheng2024zest, lopes2025matswap, zeng2024rgb, cheng2025marble}, which serve as the closest baselines for our newly defined \emph{material coating} task.
We additionally compare against two classical Photoshop operations: \emph{Blend If}~\cite{photoshopcafe-blendif} and \emph{Color Blend Mode}. Although simple and heuristic, they preserve substrate shadows and highlights well, making them strong baselines for coating effects.

Figure~\ref{fig:comparisons} shows a qualitative comparison using real images. MARBLE~\cite{cheng2025marble} is omitted as it lacks mask support (see Appendix). \ourmethod{} is the only approach that produces realistic, coherent coatings that follow the specified material properties while preserving substrate details.

Our quantitative evaluation is presented in Table~\ref{tab:metrics_results}. We measure PSNR between the results of each method and the GT image from a held-out validation split of \ourdataset{}, containing unseen objects, materials and environment maps. Since the coating task does not explicitly specify how patterned albedo textures should be mapped onto the target geometry, multiple geometrically plausible texture wrappings are possible as valid results. We therefore report PSNR separately for coatings with \emph{Uniform Albedos} (where no pattern is present), \emph{Pattern Albedos} (where mappings may differ), and the overall PSNR mean (\emph{All}).

As shown in Table~\ref{tab:metrics_results}, \ourmethod{} consistently outperforms all baselines across all evaluation subsets, highlighting its ability to generalize across material properties. 
The second-best method is the ablated version of our method, which we describe in more detail in Section~\ref{sec:ablation}.
An important observation is that simple Photoshop blending operations outperform the more sophisticated learning-based material transfer methods on this task. This strongly supports our central claim that \emph{Material Coating} is fundamentally different from \emph{Material Transfer}.

\begin{table}[ht]
    \centering
    \caption{Quantitative comparison of Material Coating, measured by PSNR. Our \ourmethod{} outperforms all baselines across all metrics.}
    \label{tab:metrics_results}
    \resizebox{\linewidth}{!}{
    \begin{tabular}{lccc}
        \toprule
        & \multicolumn{3}{c}{\textbf{Image PSNR $\uparrow$}} \\
        \cmidrule(lr){2-4}
        Method & All & Uniform Albedos & Pattern Albedos \\
        \midrule
        Marble & 19.70 & 19.89 & 19.57 \\
        ZeST & 20.51 & 21.33 & 19.93 \\
        MatSwap & 21.14 & 21.25 & 21.07 \\
        RGB$\leftrightarrow$X & 18.67 & 19.00 & 18.44 \\
        Material Fusion & 23.19 & 25.68 & 21.44 \\
        Photoshop ``Blend If'' & 21.94 & 22.48 & 21.56 \\
        Photoshop Color Blend & 23.81 & 26.35 & 22.03 \\
        Ours (ConcatAlbedo) & 24.33 & 27.19 & 22.32 \\
        Ours (No Cyclic Loss) & 24.34 & 27.56 & 22.07 \\
        Ours & \textbf{24.99} & \textbf{28.00} & \textbf{22.88} \\
        \bottomrule
    \end{tabular}
    }
\end{table}

\subsection{Ablation Study}
\label{sec:ablation}

We evaluate the importance of the proposed VICL/triptych formulation and cycle-consistent supervision by comparing against a simpler baseline, ConcatAlbedo. This baseline uses the same backbone (\textsc{Flux Fill}) but demonstrates the difficulty of geometry-aware mapping under a different architecture: we pass each of the three images to the VAE separately and concatenate the resulting latents along the channel dimension. To accommodate the expanded channel dimension, we introduce a new linear projection layer (\texttt{x\_embedder} in \textsc{Flux Fill}), copy the original weights, and zero out the new weights corresponding to the albedo latents. The LoRA is trained on this new \texttt{x\_embedder} as well as all other layers trained in the triptych method.

\begin{figure}[t]
    \centering
    \includegraphics[width=\linewidth]{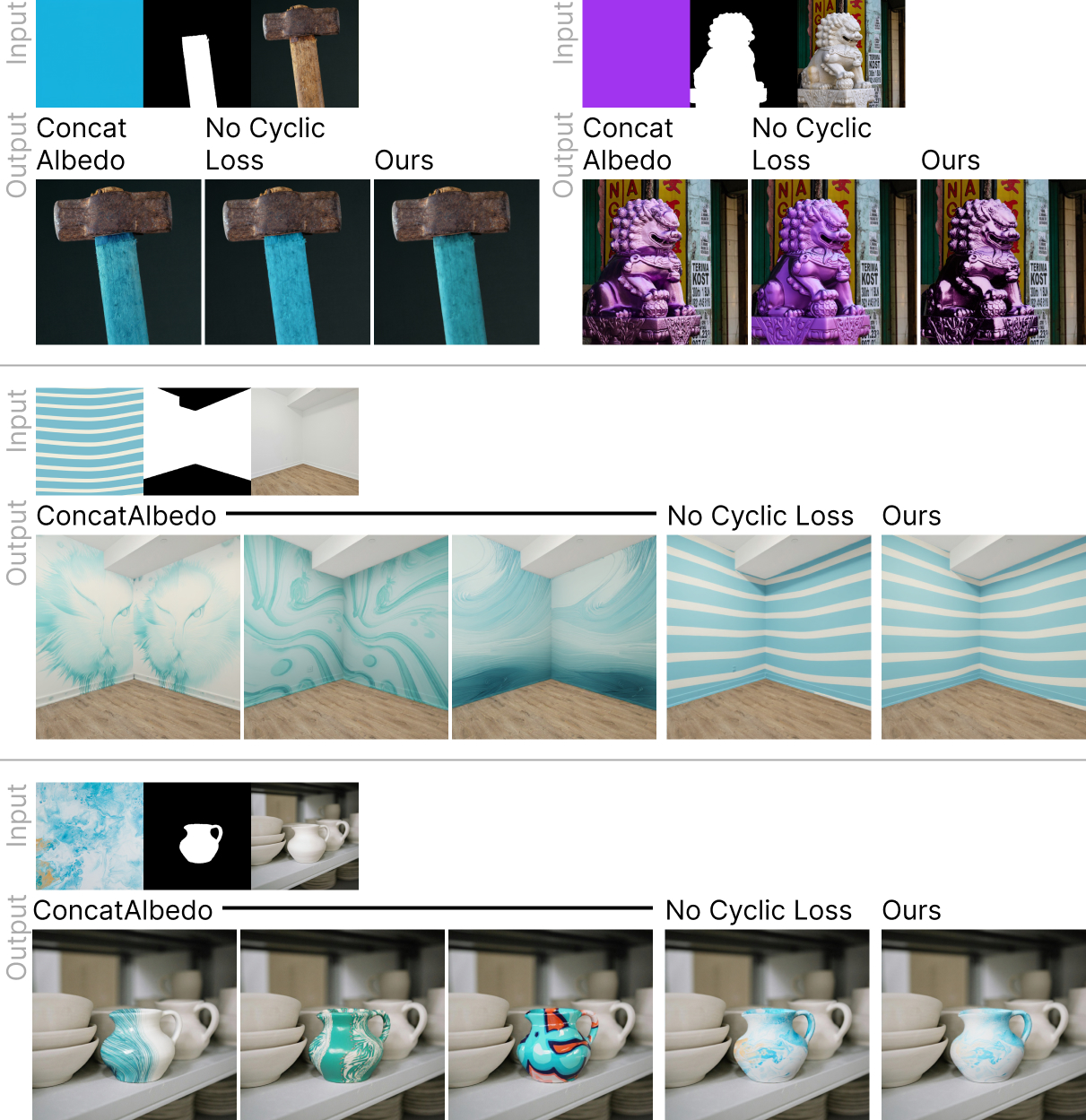}
    \caption{\textbf{Ablation study.} Latent albedo concatenation handles uniform albedo coatings (top row), but fails when a more complex geometry-aware wrapping is required. We show multiple ablation results to illustrate how this baseline fails to faithfully wrap the input albedo onto the geometry. In contrast, our proposed design plausibly maps textured coatings onto the object geometry. The no cyclic loss ablation performs well qualitatively, but yields a small but consistent drop in PSNR across all evaluation subsets.}
    \Description{Three blocks, each showing the input triptych of albedo, mask and source image above outputs labelled Concat Albedo, No Cyclic Loss and Ours. The first block holds two compact examples: a hammer whose handle is coated cyan, and a stone guardian lion coated purple. In the second block a striped teal albedo is applied to the walls of a room corner: the Concat Albedo variants produce swirling marbled smears with no recognizable stripes, while No Cyclic Loss and Ours both wrap straight horizontal stripes correctly around the corner. In the third block a blue watercolor albedo is applied to a white jug on a shelf: the Concat Albedo variants invent bold teal and orange patterns, while Ours applies a soft blue wash that preserves the jug's shading and handle.}
    \label{fig:ablation_study}
\end{figure}

This variant captures material behavior but fails at albedo wrapping (Figure~\ref{fig:ablation_study}). We attribute this to the linear projection merging each substrate patch with its spatially aligned albedo patch: an undesirable coupling, since in general no correspondence exists between a coordinate in the unwrapped albedo and the same coordinate in the input image. While ConcatAlbedo could in principle disentangle the two after the projection and wrap the albedo using self-attention, we show empirically that it fails to do so. A concatenation baseline trained on screen-space-projected albedos would sidestep this issue, but would then require projected textures from the user and be unable to perform albedo wrapping itself. By training ConcatAlbedo on the same canonical tangent-space albedos as our main model, our ablation demonstrates that without the proposed triptych formulation and loss, the model fails to learn the required object-space mapping.

In contrast, the triptych formulation explicitly exposes the relationship between the substrate image, the canonical albedo, and the coated result. Combined with cycle-consistent supervision, this encourages the model to infer object-space correspondences rather than relying on screen-space alignment. Figure~\ref{fig:ablation_study} illustrates that this significantly improves geometry-aware coating behavior. Finally, we also evaluated training the same architecture without the cyclic loss. While this ablation performs well qualitatively (Figure~\ref{fig:ablation_study}), evaluating our metrics over a large set of images revealed a small but consistent improvement in PSNR across all comparison subsets (all results, uniform albedos and textured albedos) when the cyclic loss is included.

\subsection{Human Evaluation}
\label{sec:human_eval}

To evaluate perceptual quality and controllability, we conducted a human evaluation on Amazon Mechanical Turk (AMT), following the protocol detailed in the Appendix.
We used a held-out subset of \ourdataset{}.

\begin{figure}
    \centering
    \includegraphics[width=1.0\linewidth]{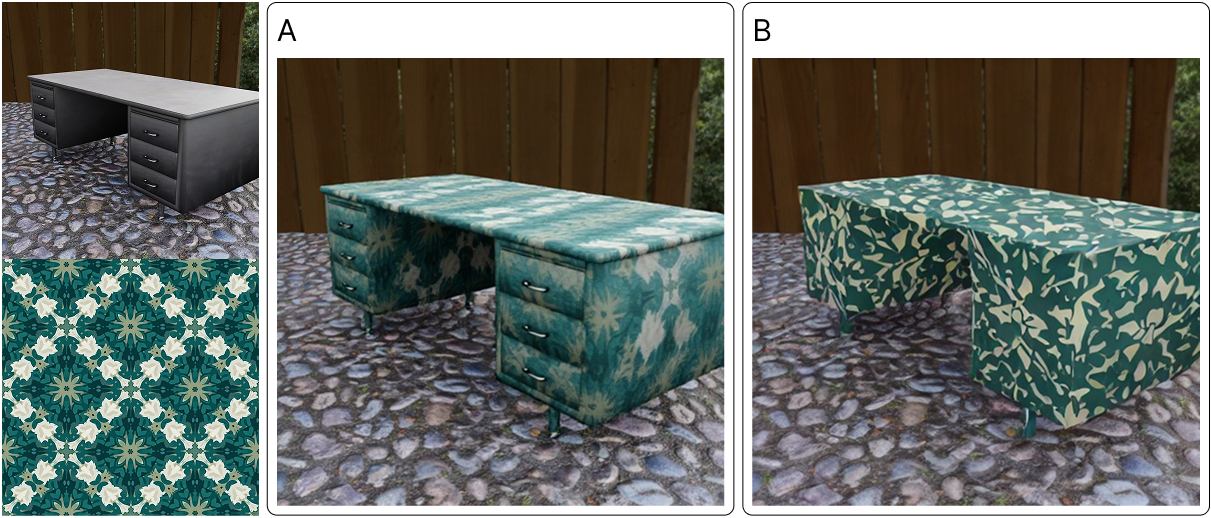}
    \caption{\textbf{Human evaluation task example.} Workers were given input and albedo images, and were tasked with choosing the most convincing material application result between \ourmethod{} and a baseline for multiple questions, one per material property. E.g. ``Which result looks like it was coated with a \textbf{metallic} coating?''}
    \Description{The layout shown to crowd workers. On the left, the source photograph of a dark metal desk on cobblestones and, beneath it, a green and cream floral albedo tile. On the right, two candidate results labelled A and B. In A the floral pattern follows the desk's own geometry, breaking around the drawer fronts and the recessed knee hole. In B the pattern is applied as though a printed cloth had been draped over the desk, hiding the drawers.}
    \label{fig:user_study_task_example}
\end{figure}

We designed an A/B test comparing \ourmethod{} against various baselines. For each task, participants were presented with the inputs (original input image and the desired albedo texture), followed by our result and a baseline result with randomized A/B positions (see Figure~\ref{fig:user_study_task_example}). For each A/B pair, participants were asked to make several independent choices indicating which result looked more like it was coated with a specific material property: thickness, roughness, metallic appearance, translucency, or exhibited more plausible mapping of the albedo pattern. 

Each Human Intelligence Task (HIT) contained 9 pairwise comparisons, where one of the pairs was intentionally duplicated as a control measure.
Each unique comparison was shown to 3 different participants, and the ``winner'' for each question was determined by a majority vote (considering only qualified responses). %

\begin{figure}[tbp]
    \centering
    \begin{subfigure}[b]{0.48\linewidth}
        \centering
        \includegraphics[width=\linewidth]{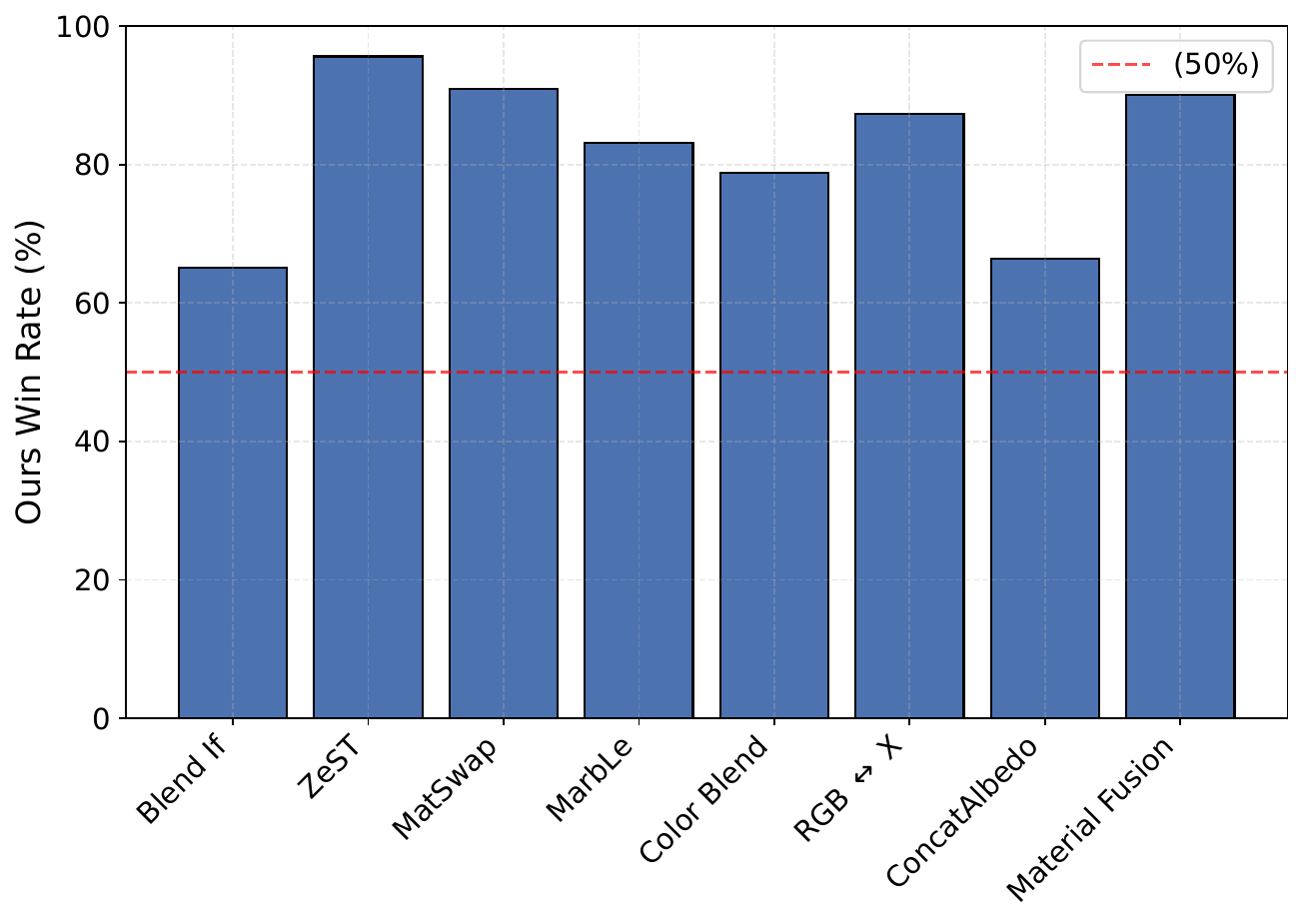}
        \label{fig:user_study_vs_method}
    \end{subfigure}
    \hfill
    \begin{subfigure}[b]{0.48\linewidth}
        \centering
        \includegraphics[width=\linewidth]{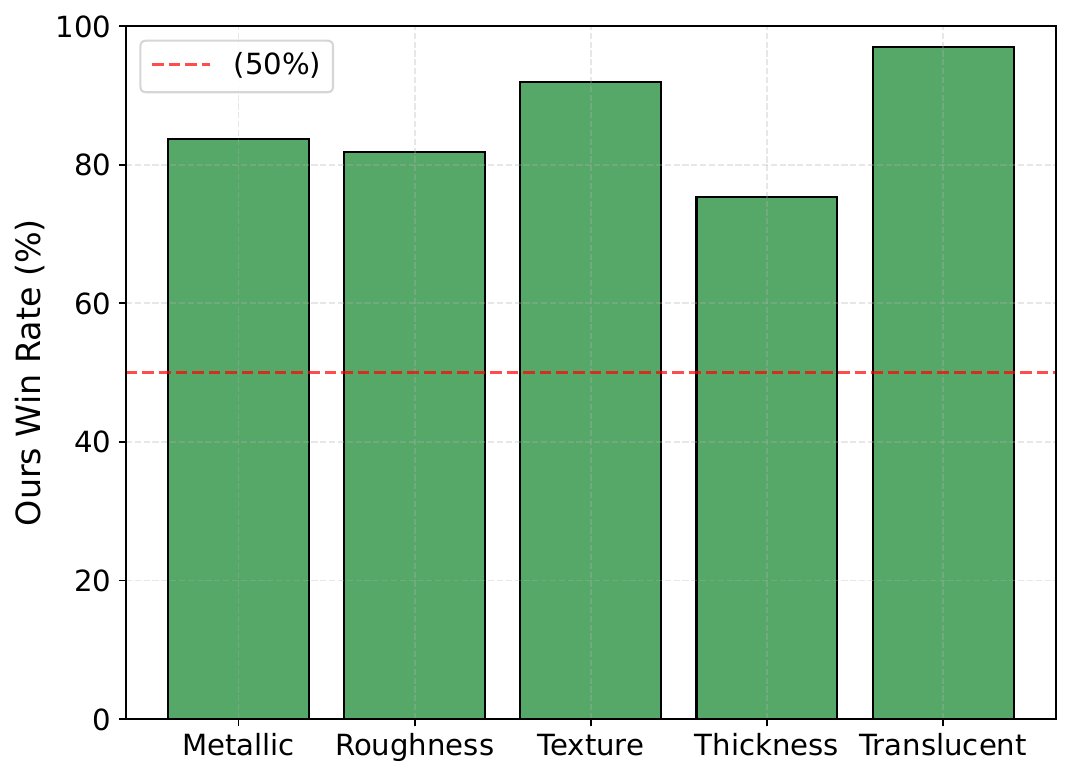}
        \label{fig:user_study_per_property}
    \end{subfigure}
    
    \begin{subfigure}[b]{1.0\linewidth}
        \centering
        \includegraphics[width=\linewidth]{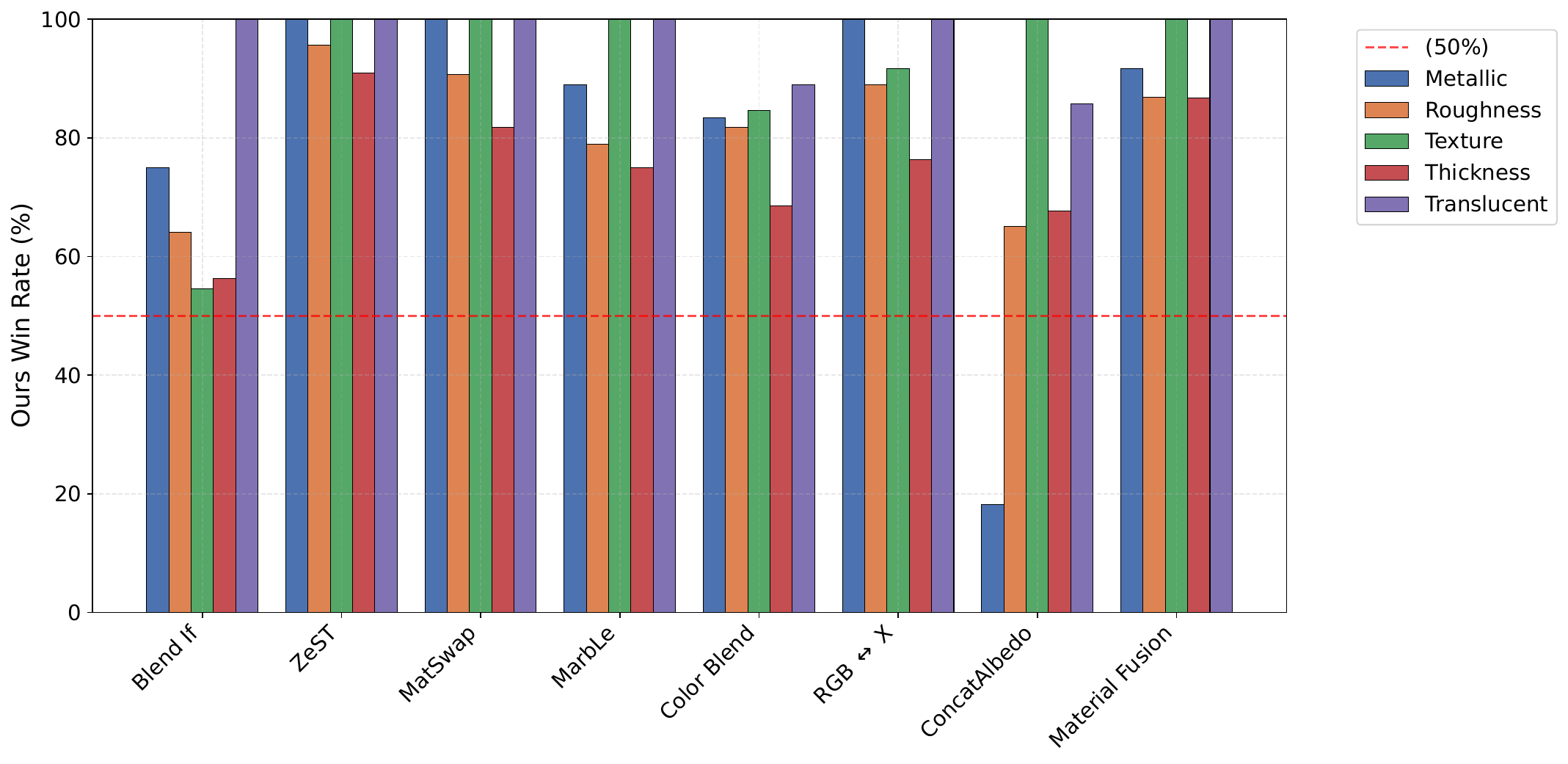}
        \label{fig:user_study_breakdown}
    \end{subfigure}
    \caption{\textbf{Human evaluation results.} Top left: Comparison against baselines, where each bar represents our aggregate win rate against that specific baseline (e.g., roughly 95\% against ZeST). Top right: Breakdown across material properties. Bottom: Win rate against each baseline, broken down by material property. ConcatAlbedo refers to our ablation baseline.}
    \Description{Three bar charts of win rates, all with a red dashed line marking 50 percent. The first plots our aggregate win rate against each baseline: Blend If 65, ZeST 96, MatSwap 91, MARBLE 83, Color Blend 79, RGB to X 87, ConcatAlbedo 66, and MaterialFusion 90 percent. The second plots the win rate per material property: Metallic 84, Roughness 82, Texture 92, Thickness 75, and Translucent 97 percent. The third breaks the win rate down by baseline and property together, with five colored bars per baseline. Every bar in the first two charts lies above the 50 percent line; in the breakdown the only bar clearly below it is ConcatAlbedo on Metallic at about 18 percent, and the lowest remaining bars are Blend If on Texture and Thickness at roughly 55 percent.}
    \label{fig:user_study_plots}
\end{figure}

Figure~\ref{fig:user_study_plots} summarizes the results. Across all evaluation questions, \ourmethod{} won \textbf{82.3\%} of comparisons against competing methods. These results confirm that our proposed approach provides better control over individual material attributes.

\begin{figure}[!ht]
    \centering
    \includegraphics[width=\linewidth]{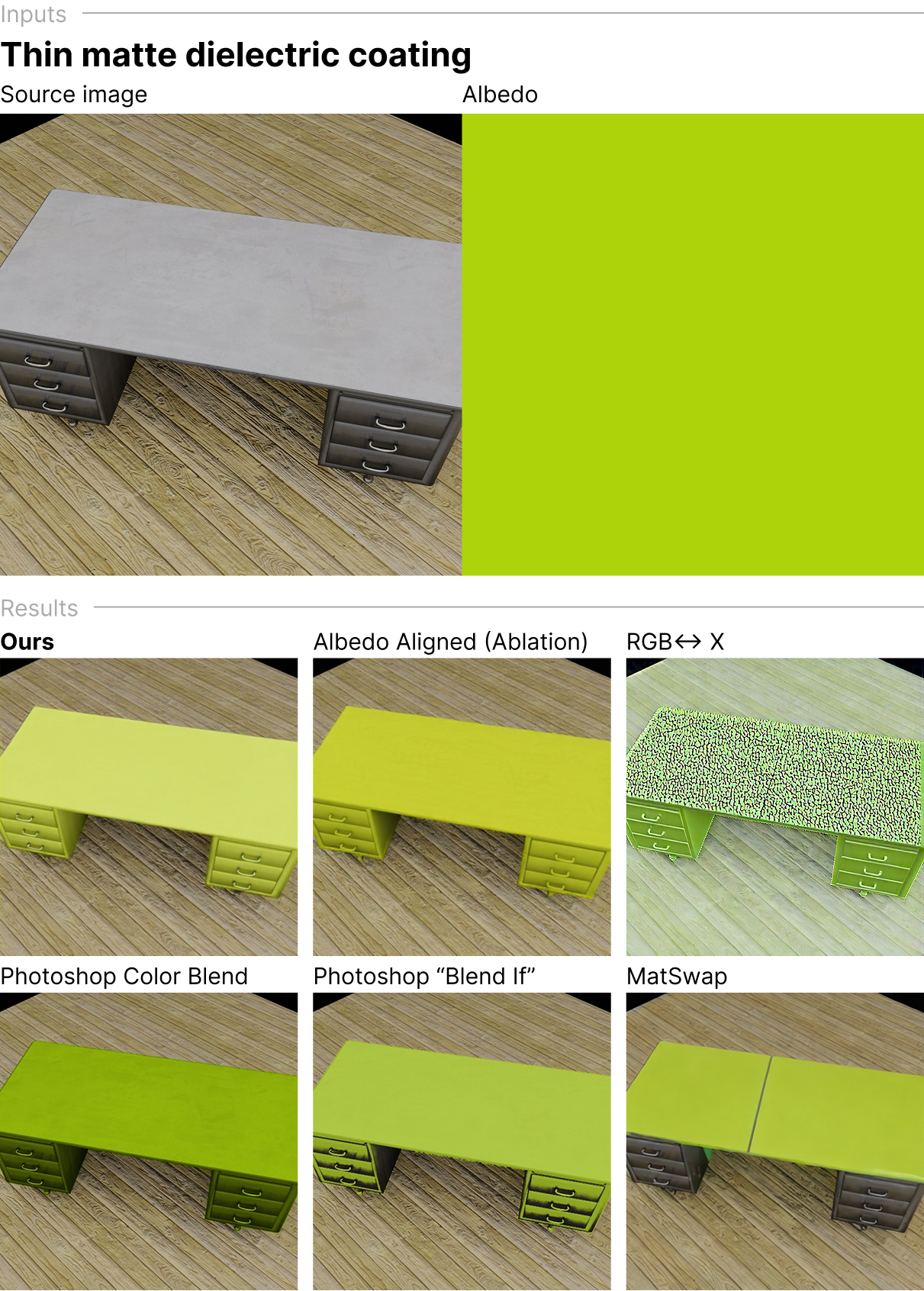}
    \caption{\textbf{Human evaluation failure case.} An example where \ourmethod{} did not achieve a clear advantage over the baselines. For this specific combination of a rough, thin, and opaque non-metallic coating, our model reached a win rate of only 40\% in both Roughness and Thickness categories.}
   \Description{At the top, the inputs: a rendered grey desk on a wooden floor and a solid yellow-green albedo, with the requested traits given as a thin matte dielectric coating. Below, six results labelled Ours, Albedo Aligned ablation, RGB to X, Photoshop Color Blend, Photoshop Blend If, and MatSwap. All six render the desk in a similar yellow-green, which is why the comparison is inconclusive; RGB to X additionally covers the tabletop with speckled artifacts, and MatSwap leaves the drawer units uncoated and dark.}
   \label{fig:user_study_bad_comparisons}
\end{figure}

For completeness, Figure~\ref{fig:user_study_bad_comparisons} shows a representative example from the user study where \ourmethod{} did not dominate. 

In this example, \ourmethod{} was preferred in only 40\% of the total votes.
We also note that our ablated variant, while incapable of wrapping albedo textures onto geometries, was judged better for uniformly colored metallic coatings.

\section{Discussion and Conclusion}

Our model generalizes well to editing material properties in real images and demonstrates implicit 3D understanding by producing plausible projections, despite being trained solely on synthetic data. However, several limitations remain.

A major limitation of our current architecture is its reliance on inpainting, which limits edits to the masked object. Consequently, the model does not produce caustics or reflective effects on the surrounding environment. Additionally, while the model autonomously determines plausible projections for patterned input, more user control for its mapping would be desirable.
Future work could address this by implementing more robust manual mapping controls.
Additionally, our method applies a single coating per inference pass; producing distinct coatings for multiple objects in a scene without requiring sequential passes with separate masks, is also left as future work.

\textbf{Conclusion.} \ourmethod{} demonstrates that controllable material coating is a tractable and impactful new editing task. By leveraging a synthetic data pipeline and geometry-aware texture mapping, we translate fine-grained material edits to real-world images. We believe this approach provides a strong foundation for future research into complex optical effects and interactive material design.

\section{Acknowledgements}

This work was supported in part by the Israel Science Foundation (grants no. 2203/24 and 1427/25).
\bibliographystyle{ACM-Reference-Format}
\bibliography{coatfusion}
\clearpage

\newpage
\appendix

\section{Appendix}

\subsection{Assets}
\label{sec:supp_assets}
\ourdataset{} was built using both downloaded and procedurally generated assets:
\begin{itemize}
    \item 116 3D models (104 for training, 12 for testing) \\ (from \texttt{polyhaven.com} \cite{polyhaven})
    \item 408 environment maps (HDRI) (from \texttt{ambientcg.com} \cite{ambientcg})
    \item 33 floor materials (from \texttt{ambientcg.com})
    \item 1500 generated ``albedo textures'' created with Flux\-dev \cite{fluxdev2025},
    including drawings, paintings, graphic design logos, graffiti, etc.
    \item $\sim$200 patterned PBR material albedo textures \\ (from \texttt{ambientcg.com} \cite{ambientcg})
\end{itemize}

\subsection{Dataset}
\label{sec:supp_dataset}
\begin{figure*}[ht]
    \centering
    \includegraphics[width=1.0\linewidth]{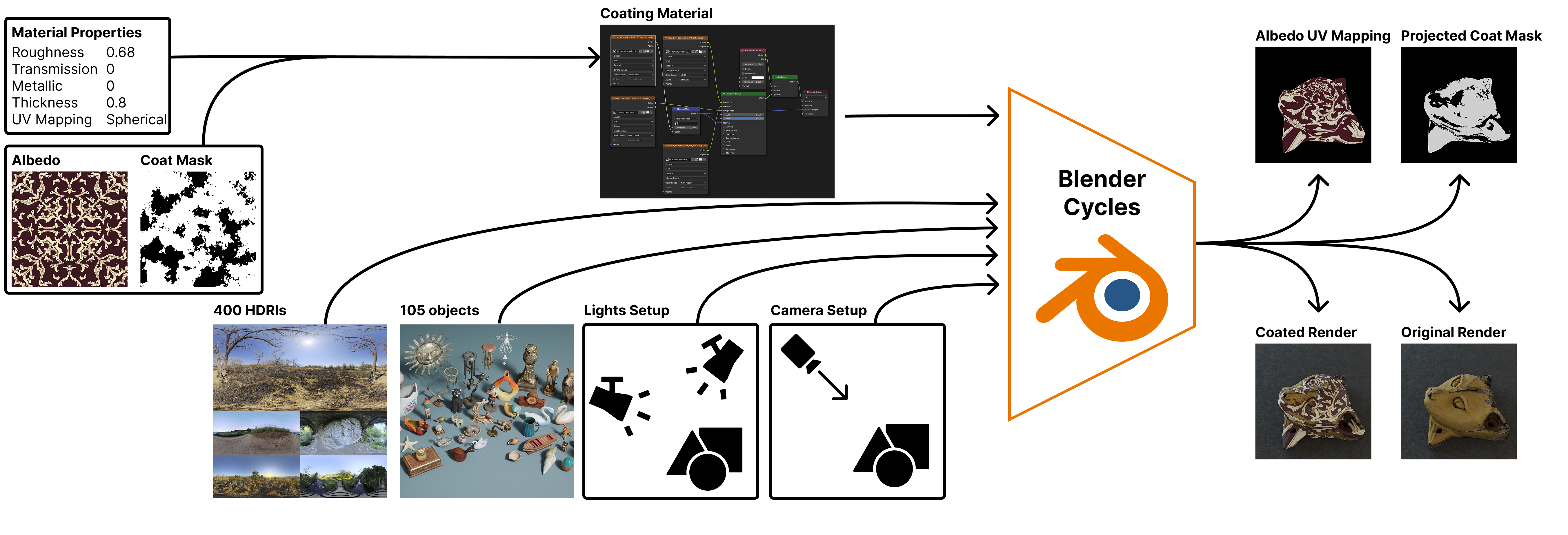}
    \caption{\textbf{Data generation}. We generate a synthetic dataset by randomly constructing scenes in Blender, rendering the original object appearance, and then generating multiple coated variants with varying coating parameters and masks (see main paper).}
    \Description{A left-to-right pipeline diagram. On the left, the randomized inputs: a material properties panel listing roughness 0.68, transmission 0, metallic 0, thickness 0.8 and spherical UV mapping; a dark red and cream damask albedo with its blotchy binary coat mask; a screenshot of the coating material's Blender shader node graph; and thumbnail strips labelled 400 HDRIs and 105 objects together with icons for the light and camera setups. Arrows from all of these feed an orange trapezoid labelled Blender Cycles, which fans out on the right to four outputs: the albedo UV mapping, the projected coat mask, the coated render of an animal skull sculpture, and the original uncoated render of the same skull.}
    \label{fig:data_generation}
\end{figure*}

The dataset generation pipeline is visualized in Figure~\ref{fig:data_generation}.

The optimal UV mapping for each mesh was manually selected by evaluating all available projection methods with patterned textures. This curation process was feasible for our set of 116 meshes and ensures that every object in the dataset utilizes its most visually plausible texture projection.

The formulas for mapping the thickness parameter (see main paper) for thickness value $x \in [0,1]$:
\begin{itemize}
    \item \textbf{Normal Strength} $1.0 - (3*x^2 - 2*x^3) * 0.7$
    \item \textbf{Opaque Albedo Blending} $1 - (1-x)^{4.5} * 0.8$
    \item \textbf{Volume Absorption Factor} $0.2 + x * 0.8$
\end{itemize}

Regarding opaque coating's intensity blending, we chose to use OkLAB for its improved perceptual lightness prediction for colors which makes color intensity manipulation more visually convincing than by using other color spaces such as HSV.

\subsection{User Study}
\label{sec:supp_user_study}
We conducted a user study to evaluate the perceptual realism of \ourmethod{} using Amazon Mechanical Turk (AMT). For this evaluation, we utilized 50 results from the DataCoat110K benchmark subset, featuring unique 3D objects, textures, and environment maps that were completely held out from the training data. To ensure the evaluations were semantically clear and informative for the workers, we applied specific constraints to the test set: metallic coatings were limited to uniform albedos, and translucent coatings were limited to uniform white albedos to prevent tinted coatings from blending with the substrate in  ways which are too subtle to notice or just unexpected.

\textbf{Task Instructions and Structure}

The study was designed as an A/B test comparing \ourmethod{} against various baseline methods. For each task, participants were presented with the original input image and the target albedo texture, alongside our result and a baseline result (with the A/B positions randomized). 

We asked multiple granular sub-questions for each image pair. Workers were tasked with determining which of the two results did a better job representing specific material properties:
\begin{itemize}
        \item Thickness
        \item Roughness
        \item Metallic appearance
        \item Translucency
        \item Mapping of the albedo on the geometry (for patterned coatings)
\end{itemize}

\textbf{Quality Control and Scoring}

Each Human Intelligence Task (HIT) contained 9 pairwise comparisons. To ensure high-quality responses, one of the comparisons within each HIT was intentionally duplicated as a control measure. If a participant's answers to the questions for the duplicated control comparison lacked consistency, their response for that entire HIT was disqualified. 

Each unique comparison was shown to 3 individual participants. The ``winner'' for each specific material property sub-question was determined by a majority vote, counting only the qualified responses.

\begin{figure}
        \centering
        \includegraphics[width=\linewidth]{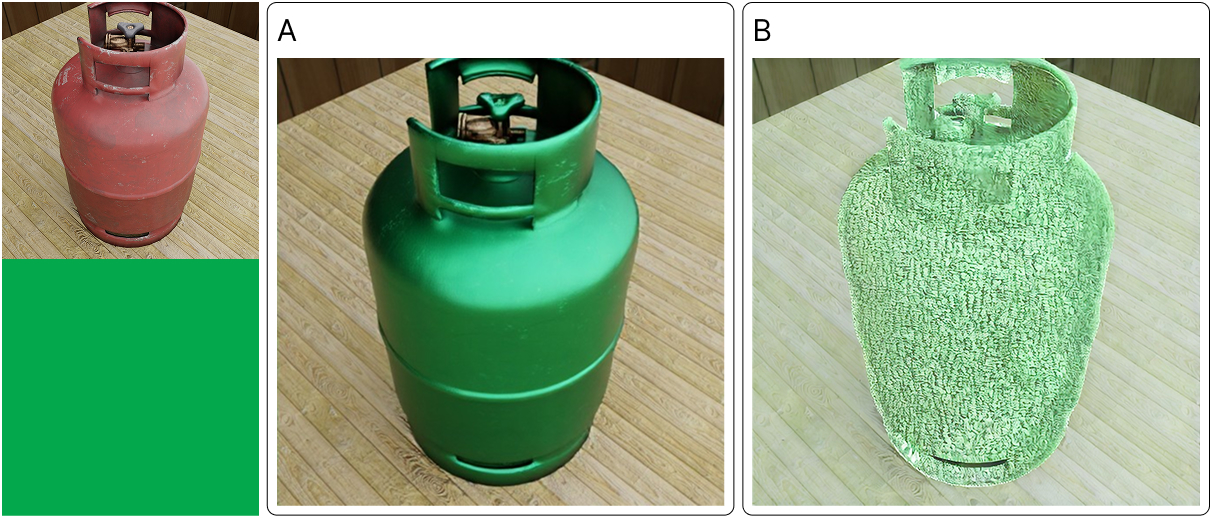}
        \caption{Example of a user study question of an object with a uniform coating.}
        \Description{A user study question. On the left, the source photograph of a faded red propane cylinder on a wooden table, with a solid green albedo below it. On the right, two candidate results labelled A and B. In A the cylinder is evenly coated in green while its valve, carry handle and welded seam stay legible. In B the green is applied as a coarse, crusted layer that swells past the cylinder's silhouette and buries the handle.}
        \label{fig:user_study_task_2}
\end{figure}

\begin{figure}
        \centering
        \includegraphics[width=\linewidth]{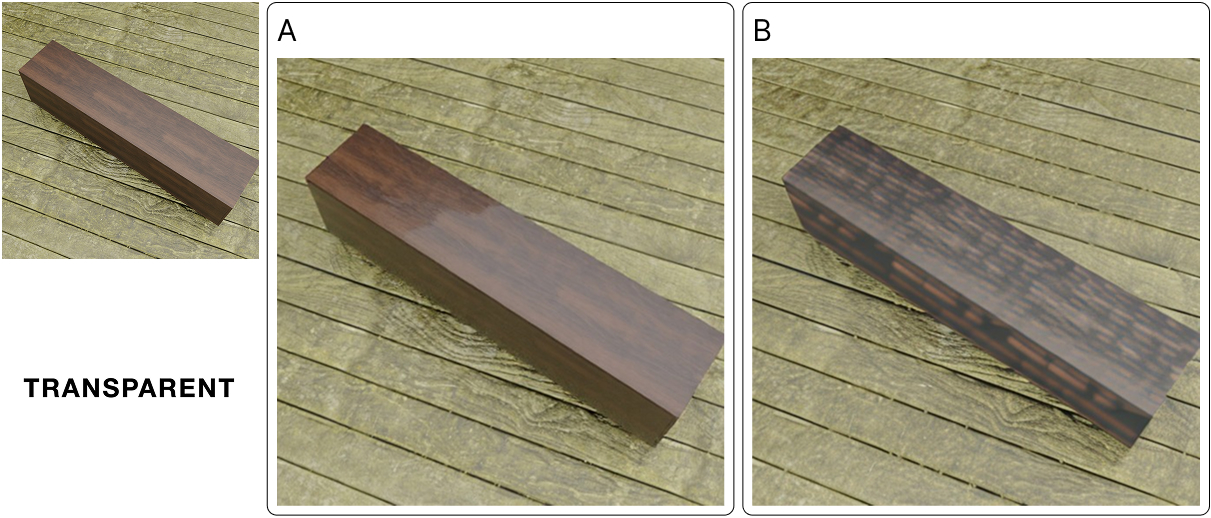}
        \caption{Example of a user study question of an object with a uniform, transparent coating.}
        \Description{A user study question. On the left, the source render of a dark wooden beam lying on a green-tinted planked floor, labelled TRANSPARENT. On the right, two candidate results labelled A and B. In A the beam is slightly darkened and glossier, with its grain still continuous along the length. In B the beam is covered in blotchy dark and orange mottling that breaks up the grain.}
        \label{fig:user_study_task_3}
\end{figure}

Below are more examples of user study tasks, where the worker is given the instructions mentioned above along with the input images and a textual representation of the material properties. For each figure, we denote the specific questions that would be asked about the material properties shown in that image.

\textbf{Figure~\ref{fig:user_study_task_2}:}
\begin{itemize}
        \item Which result looks like it was coated with a \textbf{metallic} coating?
        \item Which result looks like it was coated with a \textbf{glossy} (highly shiny) coating?
        \item Which result looks like it was coated with a \textbf{thick} coating?
\end{itemize}

\textbf{Figure~\ref{fig:user_study_task_3}:}
\begin{itemize}
        \item Which result looks like it was coated with a \textbf{translucent} coating?
        \item Which result looks like it was coated with a \textbf{glossy} (highly shiny) coating?
\end{itemize}

\subsection{Qualitative Comparison}
\label{sec:qualitative}

\begin{figure}
        \centering
        \includegraphics[width=\linewidth]{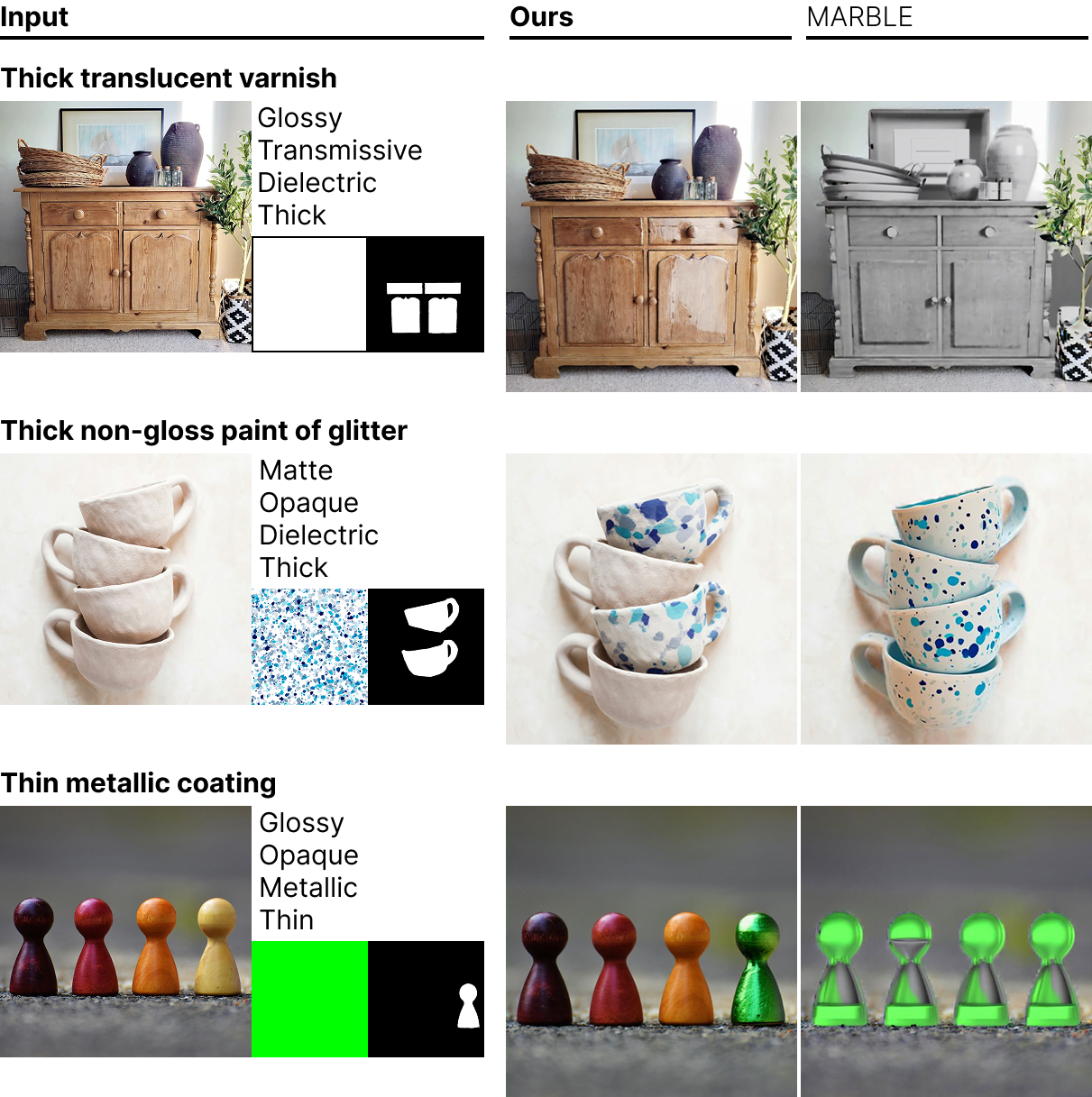}
        \caption{Comparison against MARBLE.}
        \Description{Three examples in rows and three columns: the input, our result, and MARBLE. Each input cell holds the source photograph, the requested traits as text, the albedo swatch and the mask. Row one applies a thick translucent varnish to the doors of a wooden sideboard; ours darkens the two door panels while MARBLE repaints the entire cabinet grey and alters the objects standing on it. Row two applies a thick matte glitter paint to a stack of four ceramic cups; ours keeps the cups' original cream color and adds blue flecks, while MARBLE tints the whole stack pale blue and enlarges the speckles. Row three applies a thin green metallic coating to the rightmost of four wooden pawns; ours coats only that pawn, while MARBLE turns all four into translucent green glass.}
        \label{fig:comparisons_supplementary}
\end{figure}

More comparisons against baselines methods can be seen at Figure~\ref{fig:comparisons_supplementary}.

\label{sec:quantitative}

\subsection{Complementary Task}
\begin{figure}
    \centering
    \includegraphics[width=\linewidth]{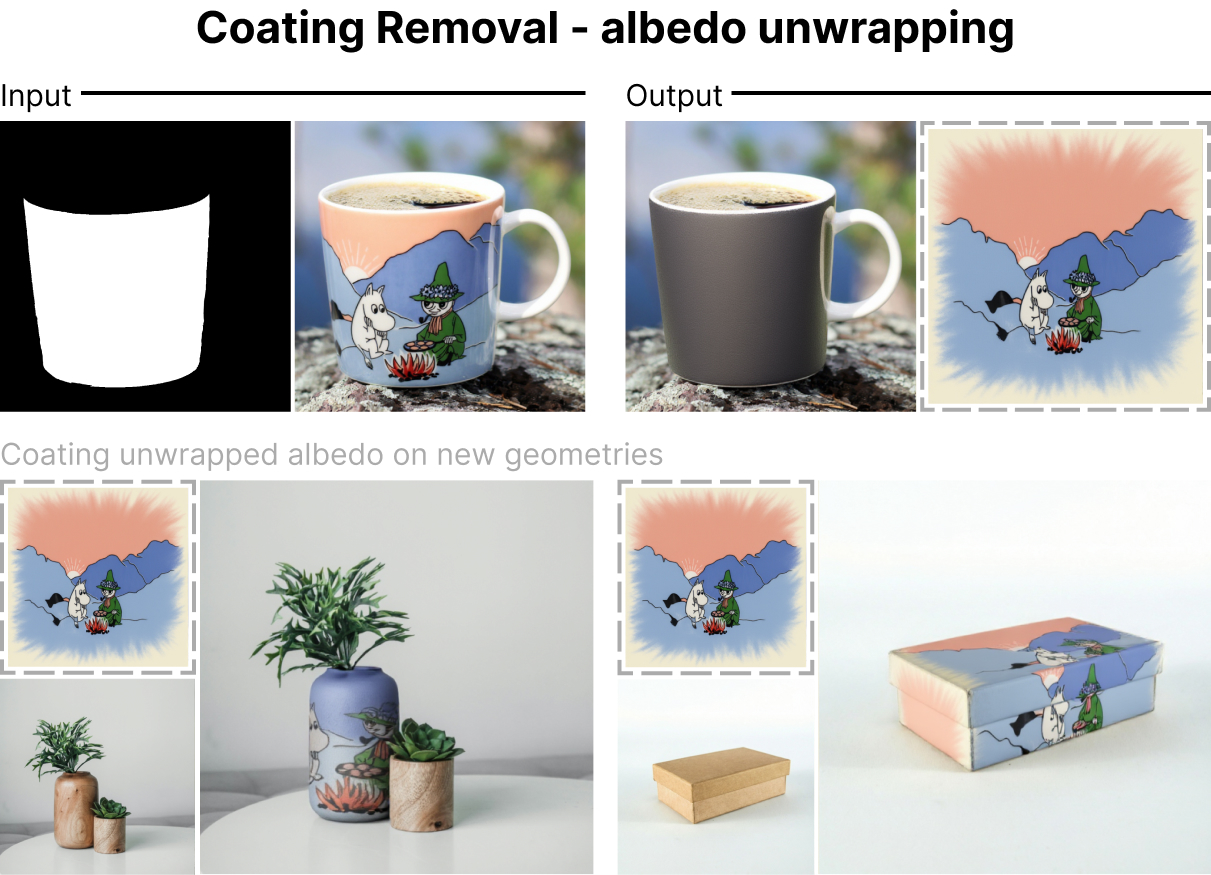}
    \caption{\textbf{Complementary Task.} Example of removing and unwrapping the albedo from the masked geometry (1st row) and coating new geometries with the unwrapped albedo (2nd row).}
    \Description{Two rows. The top row, coating removal, shows an input mask and a photograph of a mug printed with a cartoon scene of two characters by a campfire under a pink sky; the outputs are the same mug stripped back to plain dark grey ceramic and the recovered flat albedo of the scene, drawn in a dashed border. The bottom row applies that recovered albedo to two new objects: a plain terracotta vase becomes a vase carrying the campfire scene wrapped around its curvature, and a plain cardboard box becomes a box with the same scene continuing across its lid and side.}
    \label{fig:complementary_tasks}
    
\end{figure}

In addition to applying new coatings, our model is designed to handle a complementary editing task: \emph{coat removal}.  
This complementary task shares the same input structure as the coating application task: an input image together with a user-provided mask that defines the region of interest.
In the \emph{coat removal} task, the goal is to remove an existing coating from the masked region, restoring the underlying object or surface as well as the coating albedo texture in tangent space. In Figure~\ref{fig:complementary_tasks} we show a painting removal from a cup, and then use its unwrapped albedo representation to coat new geometries.

\subsection{Training Details}
\label{sec:supp_training_details}

\begin{figure*}[ht]
    \centering
    \includegraphics[width=1.0\linewidth]{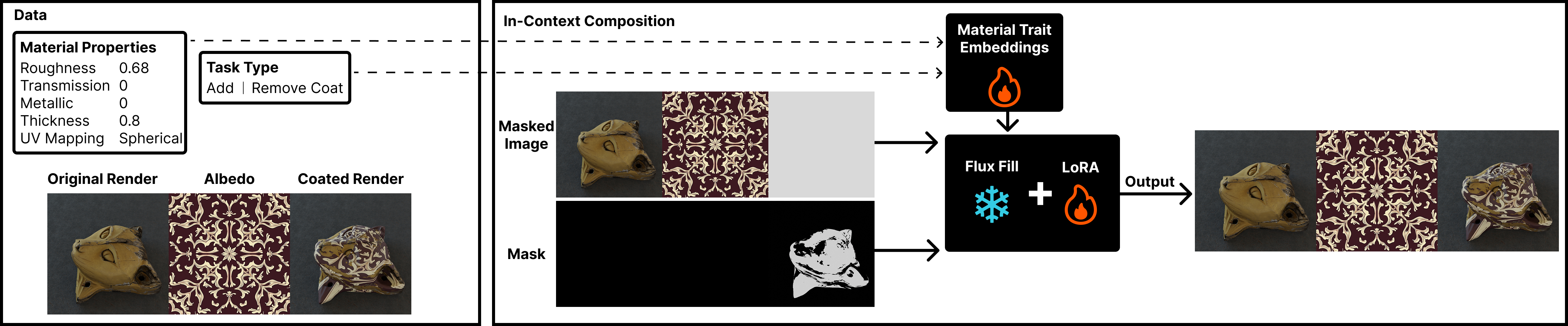}
    \caption{\textbf{Training and architecture}. Given an input image, albedo texture, user mask, and material traits, \ourmethod{} generates a coated result using a triptych-based diffusion formulation. During training, decomposition and composition are jointly optimized through cycle-consistent supervision.}
    \Description{A two-panel diagram. The left panel, Data, lists the material properties, roughness 0.68, transmission 0, metallic 0, thickness 0.8, spherical UV mapping, together with a task type of add or remove coat, and shows the training triptych: the original render of an animal skull sculpture, a dark red and cream damask albedo, and the coated render. The right panel, In-Context Composition, feeds the masked image and its binary mask into a block labelled Flux Fill with LoRA, marked with a snowflake for the frozen backbone and a flame for the trained adapter, together with material trait embeddings that are also marked as trained. An arrow labelled Output leads to the completed triptych.}
    \label{fig:architecture}
\end{figure*}

We trained \ourmethod{} using a dataset of 110K synthetic triptychs generated from our DataCoat110K dataset. Our training objective follows the flow-matching formulation, optimized over a dual-track strategy to enforce geometric and material consistency via a cyclic loss.

\begin{itemize}
    \item \textbf{Triptych Configuration:} Training samples are structured as $1536 \times 512$ triptychs. Each sample contains the input object, the target material albedo, and the resulting coated object. By concatenating these into a single image, we leverage the model's self-attention to learn geometry-aware wrapping and spatial correspondence.
    
    \item \textbf{Training tracks:} We utilize two complementary tracks to enable the cyclic loss:
        \begin{itemize}
            \item \textit{Coating Track ($F_C$):} The model learns to apply a material albedo onto a 3D substrate.
            \item \textit{Decomposition Track ($F_D$):} The model learns to invert the process, extracting the 2D albedo from a 3D coated object.
        \end{itemize}
    We calculate one track per step due to memory constraints, alternating between them.

    \item \textbf{Parameters:} 
        \begin{itemize}
            \item \textbf{Hardware \& Steps:} The initial 30,000 steps ran on 3 H200 GPUs for 26 hours at 256x256 resolution (total batch size 36). We then continued for 5,000 steps (12 hours) at 512x512 resolution on 4 H200 GPUs (total batch size 32). This two-stage schedule speeds up the training progress while maintaining the model's ability to generate results at higher resolutions.
            \item \textbf{Optimizer:} AdamW with a constant learning rate of $1\times 10^{-4}$ for both runs.
            \item \textbf{LoRA:} We trained only a LoRA of rank 128, keeping the \textsc{Flux Fill} backbone frozen.
        \end{itemize}

    \item \textbf{Trait Embeddings:} Parametric controls (thickness, roughness, metalness, transmission) are injected as global trait embeddings:
        \begin{itemize}
            \item \textbf{Continuous Traits:} Thickness and roughness are \\ mapped from $[0,1]$ to a learned embedding space.
            \item \textbf{Discrete Traits:} For binary traits we train distinct ``On/Off'' embeddings: metallic vs.\ dielectric, and transmissive vs.\ opaque.
            \item \textbf{Initialization:} Each value embedding is initialized from the CLIP text-encoder tokens of the corresponding semantic terms (e.g., `varnish', `metallic', `matte') to align the latent space with the model's prior knowledge.
        \end{itemize}
\end{itemize}

\subsection{More results}

\begin{figure*}[ht]
    \centering
    \includegraphics[width=1.0\linewidth]{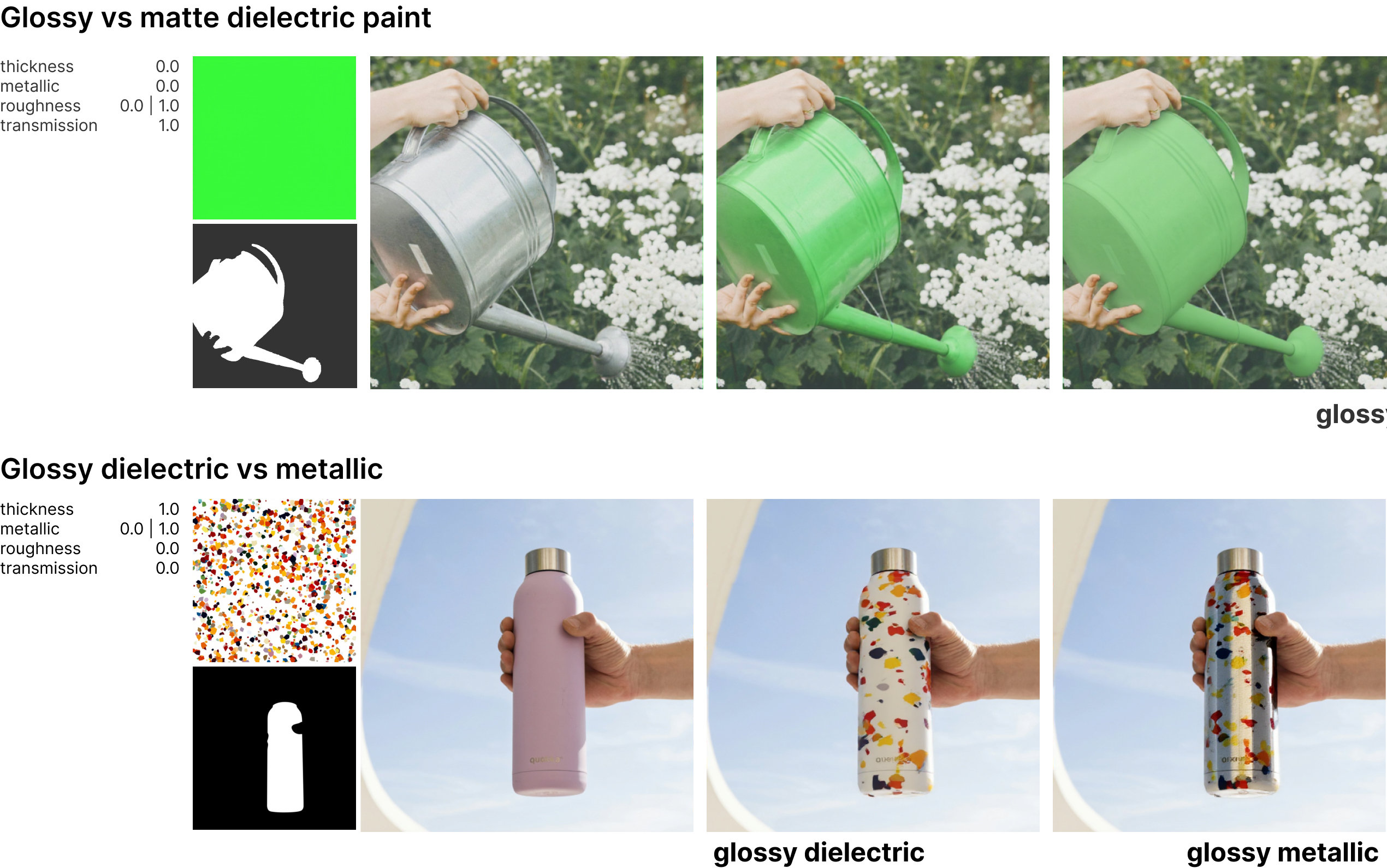}
    \caption{\textbf{Additional Results.} Our method generalizes to various objects and materials. \textbf{Top:} A watering can coated with glossy and matte green finishes. \textbf{Bottom:} A water bottle featuring patterned coatings with varying material traits.}
    \Description{Two blocks, each showing the material parameters and albedo on the left and the source photograph beside two results. In the first block, glossy versus matte dielectric paint, a galvanized metal watering can held over white flowers is coated in solid green: the glossy result keeps bright specular highlights along the rim and spout, while the matte result is uniformly flat. In the second block, glossy dielectric versus metallic, a plain pink water bottle held against the sky is coated with a confetti speckle pattern: the dielectric result reads as painted enamel, while the metallic result gives the same speckles a reflective sheen and a darker, mirror-like ground.}
    \label{fig:more_results_supp}
\end{figure*}

We provide additional qualitative results in Figure~\ref{fig:more_results_supp}, showcasing our model's ability to handle more geometries and material transitions. For instance, a metallic watering can is successfully coated with both metallic and matte green finishes, while maintaining the underlying specular highlights and geometric details. Similarly, we demonstrate texture wrapping on a water bottle, applying dotted patterns with varying metallic and non-metallic properties.

\end{document}